\documentclass[prd,amsfonts,amssymb,noshowpacs,nofootinbib,preprintnumbers]{revtex4}
\usepackage{amsmath,graphicx,hyperref,epsfig}

\begin{document}

\preprint{YITP-11-61}
\title{Large and strong scale dependent bispectrum in single field inflation from a sharp feature in the mass}
\author{Frederico Arroja$^{1}$\footnote{arrojaf@ewha.ac.kr}, Antonio Enea Romano$^{2}$\footnote{aer@phys.ntu.edu.tw} and Misao Sasaki$^{3}$\footnote{misao@yukawa.kyoto-u.ac.jp}}
\affiliation{
{}$^{1}$Institute for the Early Universe, Ewha Womans University, Seoul 120-750, Republic of Korea\\
{}$^{2}$Leung Center for Cosmology and Particle Astrophysics, National Taiwan University, Taipei 106, Taiwan\\
{}$^{3}$Yukawa Institute for Theoretical Physics, Kyoto University, Kyoto 606-8502, Japan;
Korea Institute for Advanced Study
207-43 Cheongnyangni 2-dong, Dongdaemun-gu, Seoul 130-722, Republic of Korea
}

\begin{abstract}
We study an inflationary model driven by a single minimally coupled standard kinetic term scalar field with a step in its mass modeled by an Heaviside step function. We present an analytical approximation for the mode function of the curvature perturbation, obtain the power spectrum analytically and compare it with the numerical result. We show that, after the scale set by the step, the spectrum contains damped oscillations that are well described by our analytical approximation. We also compute the dominant contribution to the bispectrum in the equilateral and the squeezed limits and find new shapes. In the equilateral and squeezed limits the bispectrum oscillates and it has a linear growth envelope towards smaller scales. The bispectrum size can be large depending on the model parameters.
\end{abstract}

%\pacs{98.80.-k, 04.30.-w, 98.80.Cq}

\date{\today}
\maketitle

%%%%%%%%%%%%%%%%%%%%%%%%%%%%%%%%%%%%%%%%%%%%%%%%%%%%%%%%%%%%%%%%%%%%
%%%%%%%%%%%%%%%%%%%%%%%%%%%%%%%%%%%%%%%%%%%%%%%%%%%%%%%%%%%%%%%%%%%%%
%%%%%%%%%%%%%%%%%%%%%%%%%%%%%%%%%%%%%%%%%%%%%%%%%%%%%%%%%%%%%%%%%%%%%
\section{Introduction}

The inflationary paradigm \cite{Starobinsky:1979ty,Kazanas:1980tx,Sato:1980yn,Guth:1980zm,Linde:1981mu,Albrecht:1982wi} solves successfully the problems of the Big Bang model and also provides us with a mechanism to generate the primordial perturbations from quantum fluctuations that are later imprinted as anisotropies in the cosmic microwave background (CMB) radiation and in the large-scale structure of the universe.
In particular, regarding the CMB power spectrum, we now know that it is nearly scale invariant \cite{Komatsu:2010fb}, mostly of adiabatic origin and it contains almost all the information about the primordial perturbations (i.e. the perturbations are nearly Gaussian). This is in agreement with the predictions of many models of inflation.

However, a small amount of non-Gaussianity is still allowed by the CMB data and recently there have been several claims of detection of non-Gaussianity of the primordial perturbations \cite{Yadav:2007yy,Xia:2010yu,Hoyle:2010ce,Enqvist:2010bg}. If the values of the non-Gaussianity parameters are of the order of magnitude claimed then NASA's satellite, Planck \cite{PLANCK}, which is already up in the sky taking data, should be able to measure these values at many standard deviations \cite{Komatsu:2001rj}.
These are exciting times for theoretical cosmologists that for the past ten years or so have been trying to construct models where large non-Gaussianity is produced.

It is well known that the simplest and most popular inflationary model, a single scalar field with standard kinetic term satisfying the slow-roll conditions and with standard initial conditions for the vacuum state of the quantum perturbation predicts a level of non-Gaussianity that is small and unobservable \cite{Maldacena:2002vr,Acquaviva:2002ud,Seery:2006vu,Seery:2008ax,Seery:2006js,Byrnes:2006vq}, even for the Planck satellite \cite{Kogo:2006kh,Desjacques:2009jb,Smidt:2010ra}. Therefore, the previously mentioned observations have the potential if confirmed to rule out a large and most popular class of models. It is good on itself to rule out many models in one go and even better that a detection of non-Gaussianity (the same applies to the tensor-to-scalar ratio) would allows us to progress significantly in our understanding of the mechanism that drove inflation in the early universe, this is because higher-order statistics contain much more information about the dynamics that the power spectrum.

Our searches for inflationary models producing large observable non-Gaussianity have been fruitful. Many possibilities have been found, for example models with non-canonical kinetic terms (like DBI-inflation, k-inflation, ghost-inflation) \cite{Creminelli:2003iq,Alishahiha:2004eh,Gruzinov:2004jx,Chen:2006nt,Huang:2006eha,Arroja:2008ga,Chen:2009bc,Arroja:2009pd,Huang:2010ab,Izumi:2010wm,Mizuno:2010ag,Burrage:2010cu}, multiple field models of inflation \cite{Dvali:2003em,Enqvist:2004ey,Lyth:2005qk,Lyth:2005fi,Alabidi:2006hg,Sasaki:2006kq,Valiviita:2006mz,Sasaki:2008uc,Naruko:2008sq,Suyama:2008nt,Byrnes:2008wi,Byrnes:2008zz,Byrnes:2008zy,Cogollo:2008bi,Rodriguez:2008hy,Gao:2008dt,Langlois:2008vk,Langlois:2008wt,Langlois:2008qf,Arroja:2008yy,Chen:2009zp,Huang:2009xa,Huang:2009vk,Byrnes:2009qy,Langlois:2009ej,Mizuno:2009cv,Mizuno:2009mv,Gao:2009at,RenauxPetel:2009sj,Cai:2009hw,Kim:2010ud,Gao:2010xk},
temporary violations of the slow-roll conditions and small departures of the initial vacuum state from the standard Bunch-Davies vacuum \cite{Chen:2006xjb,Chen:2008wn,Hotchkiss:2009pj,Hannestad:2009yx,Flauger:2010ja,Chen:2010bk,Takamizu:2010xy,Agullo:2010ws,Ganc:2011dy}. For recent reviews about these mechanisms to produce non-Gaussian perturbations see \cite{Koyama:2010xj,Chen:2010xk,Tanaka:2010km,Byrnes:2010em,Wands:2010af}.

Both at the bispectrum level (three-point correlation function) and the trispectrum level (four-point correlation function) many observationally distinct shapes of these higher-order correlations have been found \cite{Babich:2004gb,Senatore:2009gt,Bartolo:2010bj,Bartolo:2010di,Senatore:2010wk}. Some of these shapes of non-Gaussianity have been constrained with CMB data \cite{Komatsu:2010fb,Fergusson:2010dm} and large-scale structure data \cite{Slosar:2008hx,Desjacques:2009jb,Smidt:2010ra,Xia:2010pe,DeBernardis:2010kc}.
Current limits on the amplitude of the bispectrum, taken from the 7-year WMAP \cite{WMAP} data at 95\% confidence level are \cite{Komatsu:2010fb}: $-10<f_{NL}^{local}<74$, $-214<f_{NL}^{equil}<266$ and $-410<f_{NL}^{ortho}<6$ for the local, equilateral and orthogonal shapes respectively.
Ref. \cite{Fergusson:2010dm} has a more recent analysis of the CMB bispectrum (see \cite{Fergusson:2010gn} for the trispectrum analysis) including many other shapes. Planck satellite will improve all these constraints significantly and will also improve existing constraints on the amplitude of the trispectrum. For some review papers on observational aspects of non-Gaussianity see for instance \cite{Liguori:2010hx,Komatsu:2010hc,Yadav:2010fz,Desjacques:2010jw,Desjacques:2010nn}.

In this paper, in order to produce large and observable non-Gaussianity, we shall study a model where the slow-roll conditions are temporarily violated. This model can be included in the class of models with ``features" (i.e. non-scale invariant models) in the potential or more generally in the field's Lagrangian \cite{Starobinsky:1992ts,Adams:2001vc,Gong:2005jr,Joy:2007na,Biswas:2010si,Nakashima:2010sa,Chen:2011zf}. There are several other reasons that motivate the study of ``feature(s)" models. For instance, it has been shown that these models can provide better fits to the power spectrum of the CMB anisotropies than the $\Lambda\mathrm{CDM}$ model \cite{Covi:2006ci,Hamann:2007pa,Joy:2008qd,Mortonson:2009qv,Hazra:2010ve}. In these models, this is achieved thanks to the introduction of new parameters (and scales) that can be tuned to coincide with well known ``glitches" $\ell\sim20-40$ in the CMB spectrum data .
These models are also of theoretical interest because they have very distinct bispectrum signatures like scale-dependence (e.g. ``ringing" and localization of $f_{NL}$). The features often have a more fundamental physical explanation, like for example they might be localized sharp features due to particle production during inflation \cite{Chung:1999ve,Romano:2008rr,Barnaby:2009dd,Barnaby:2010ke},  due to a duality cascade in brane inflation \cite{Bean:2008na}, periodic features due to the production of instantons in axion monodromy inflation \cite{Silverstein:2008sg,McAllister:2008hb,Flauger:2009ab} or phase transitions \cite{Adams:1997de}. The study of these more realistic scenarios might eventually allows us to identify the underlying microscopic theory of inflation.

In particular, in this paper we will study a model where the mass of the inflaton field suddenly changes. We will describe this using a toy model where the change is approximated by a Heaviside step function. For example, this might be a toy model for a first order phase transition. We will compute the power spectrum and the bispectrum of the curvature perturbation and show that its amplitude can be large and that its shape is new with very distinctive features.

This paper is divided into the following sections. In section \ref{sec:model}, we introduce the model and give the analytical background solution under some approximations. In section \ref{sec:pert} we discuss linear perturbations, present an analytical approximation for the mode function of the primordial curvature perturbation and finally calculate the analytical power spectrum and compare it with the numerical result.
In section \ref{sec:bispectrum}, we compute the dominant contribution to the bispectrum of the curvature perturbation in two interesting limits, namely the equilateral limit and the squeezed limit. Section \ref{sec:conclusion} is devoted to the conclusion.

%%%%%%%%%%%%%%%%%%%%%%%%%%%%%%%%%%%%%%%%%%%%%%%%%%%%%%%%%%%%%%%%%%%%%%%
%%%%%%%%%%%%%%%%%%%%%%%%%%%%%%%%%%%%%%%%%%%%%%%%%%%%%%%%%%%%%%%%%%%%%%%%
%%%%%%%%%%%%%%%%%%%%%%%%%%%%%%%%%%%%%%%%%%%%%%%%%%%%%%%%%%%%%%%%%%%%%%%%%
\section{The model\label{sec:model}}

We will consider a canonical minimally coupled single scalar field model with the potential given by
\begin{eqnarray}
V(\phi)=\left\{
                \begin{array}{lr}
                   V_0+\frac{1}{2}m_\phi^2\phi^2 & : \phi > \phi_0\\
                   V_{0a}+\frac{1}{2}m_\phi^2(1+A)\phi^2 & : \phi < \phi_0
                \end{array}
         \right.\label{potential}
\end{eqnarray}
where $\phi$ is the scalar field, $m_\phi$ its mass, $V_0$ and $V_{0a}$ are the vacuum energies before and after the transition field value $\phi_0$ respectively. $A$ is a constant amplitude and a parameter of the model. We restrict the allowed range of $A$ as $A>-1$.
Requiring continuity of the potential across $\phi_0$ implies $V_{0a}=V_0-1/2m_\phi^2A\phi_0^2$, where $\phi_0=\phi(t_0)$ and $t_0$ is the transition time. We will assume that this change in vacuum energy is small and we shall neglect it.
In the next section, we will consider perturbations. We will perform the calculation in the comoving gauge therefore the background value of $\phi$, i.e. $\phi_0$, fully determines when the transition happen even in the perturbed spacetime.

We are interested in flat, homogeneous, and isotropic
Friedmann-Lema\^{\i}tre-Robertson-Walker background universes
described by the line element
\begin{equation}
ds^2=-dt^2+a^2(t)\delta_{ij}dx^idx^j\,, \label{FRW}
\end{equation}
where $a(t)$ is the scale factor. The Friedmann equation reads
\begin{equation}
H^2=\frac{1}{3M^2_{Pl}}\left(\frac{1}{2}\dot\phi^2+V(\phi)\right),
\end{equation}
where the Hubble rate is $H=\dot{a}/a$, dot denotes derivative with respect to cosmic time $t$ and $M_{Pl}$ denotes the reduced Planck mass. We will assume that $V(\phi)$ is dominated by the vacuum energy $V_0$, therefore the previous equation simplifies to
\begin{equation}
H^2\approx\frac{V_0}{3M^2_{Pl}}.
\end{equation}

The equation of motion for the scalar field reads
\begin{equation}
\ddot \phi(t) +3H\dot \phi(t)+m^2_\phi\phi(t)\left[1+A\theta(\phi_0-\phi)\right]=0,\label{Klein-Gordon}
\end{equation}
where $\theta(\phi)$ denotes the Heaviside step function.

The slow roll parameters are defined as
\begin{equation}
\epsilon=-\frac{\dot H}{H^2}=\frac{\dot\phi^2}{2H^2M^2_{Pl}}, \quad \eta=\frac{\dot \epsilon}{\epsilon H}=2\left(\frac{\ddot\phi}{\dot\phi H}+\epsilon\right).
\end{equation}

With the assumption of a vacuum dominated inflationary model, the solution of Eq. (\ref{Klein-Gordon}) before $t_0$ is
\begin{equation}
\phi_b(t)=\phi_{0b}U_b^+(t)+\tilde\phi_{0b}U_b^-(t),
\end{equation}
where $\phi_{0b}$ and $\tilde\phi_{0b}$ are the initial conditions, which we choose such that $\tilde\phi_{0b}$ is equal to zero. $U_b^\pm(t)$ are defined as
\begin{equation}
U_b^\pm(t)=e^{\lambda_b^\pm Ht},
\end{equation}
with $\lambda_b^\pm$ defined as
\begin{equation}
\lambda_b^\pm=-\frac{3}{2}\left(1\mp\sqrt{1-\frac{4}{9}\mu^2}\right),
\end{equation}
where $\mu^2$ is $\mu^2=\frac{m_\phi^2}{H^2}$.
If $\mu^2$ is small then $\lambda_b^+\approx-\frac{\mu^2}{3}$ and $\lambda_b^-\approx-3\left(1-\frac{\mu^2}{9}\right)$.
After $t_0$ the solution of Eq. (\ref{Klein-Gordon}) is
\begin{equation}
\phi_a(t)=\phi_{0a}U_a^+(t)+\tilde\phi_{0a}U_a^-(t),
\end{equation}
where $\phi_{0a}$ and $\tilde\phi_{0a}$ are integration constants and the functions $U_a^\pm(t)$ are defined as
\begin{equation}
U_a^\pm(t)=e^{\lambda_a^\pm Ht},
\end{equation}
with $\lambda_a^\pm$ defined as
\begin{equation}
\lambda_a^\pm=-\frac{3}{2}\left(1\mp\sqrt{1-\frac{4}{9}\mu_a^2}\right),
\end{equation}
where $\mu_a^2$ is $\mu_a^2=\mu^2(1+A)$. $U^+(t)$ describes the slow-roll solution while $U^-(t)$ the rapidly decaying solution.

Imposing the continuity of $\phi$ and $\dot\phi$ across $t_0$ determines the integration constants $\phi_{0a}$ and $\tilde\phi_{0a}$ as
\begin{equation}
\phi_{0a}=\phi_{0b}\frac{\lambda^-_a-\lambda^+_b}{\lambda^-_a-\lambda^+_a}e^{Ht_0\left(\lambda^+_b-\lambda^+_a\right)},
\quad
\tilde\phi_{0a}=\phi_{0b}\frac{\lambda^+_b-\lambda^+_a}{\lambda^-_a-\lambda^+_a}e^{Ht_0\left(\lambda^+_b-\lambda^-_a\right)}.
\end{equation}

In the plots of the following sections, we will make the following choice for the parameters of the model
\begin{equation}
m_\phi=6\times10^{-9}M_{Pl}, \quad H=2\times10^{-7}M_{Pl},  \quad \phi_{0b}=10M_{Pl}.
\end{equation}
We should note that the use of trans-Planckian field values might not be well justified in the context of the effective field theory, where higher dimensional operators are expected to destroy the flatness of the potential.

\begin{figure}[t]
\centering
 \scalebox{.28}
 {\rotatebox{0}{
    \includegraphics*{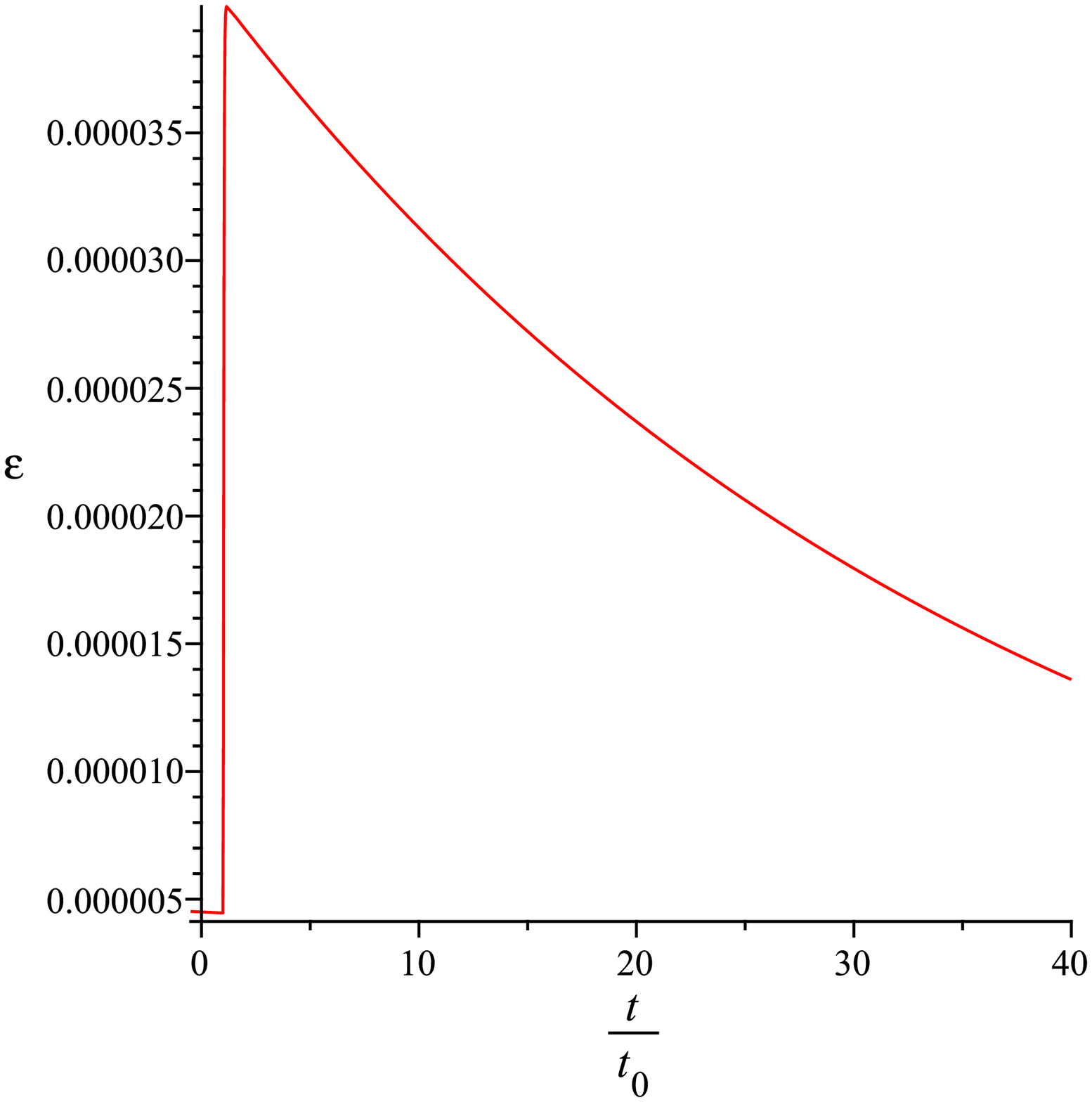}
    \includegraphics*{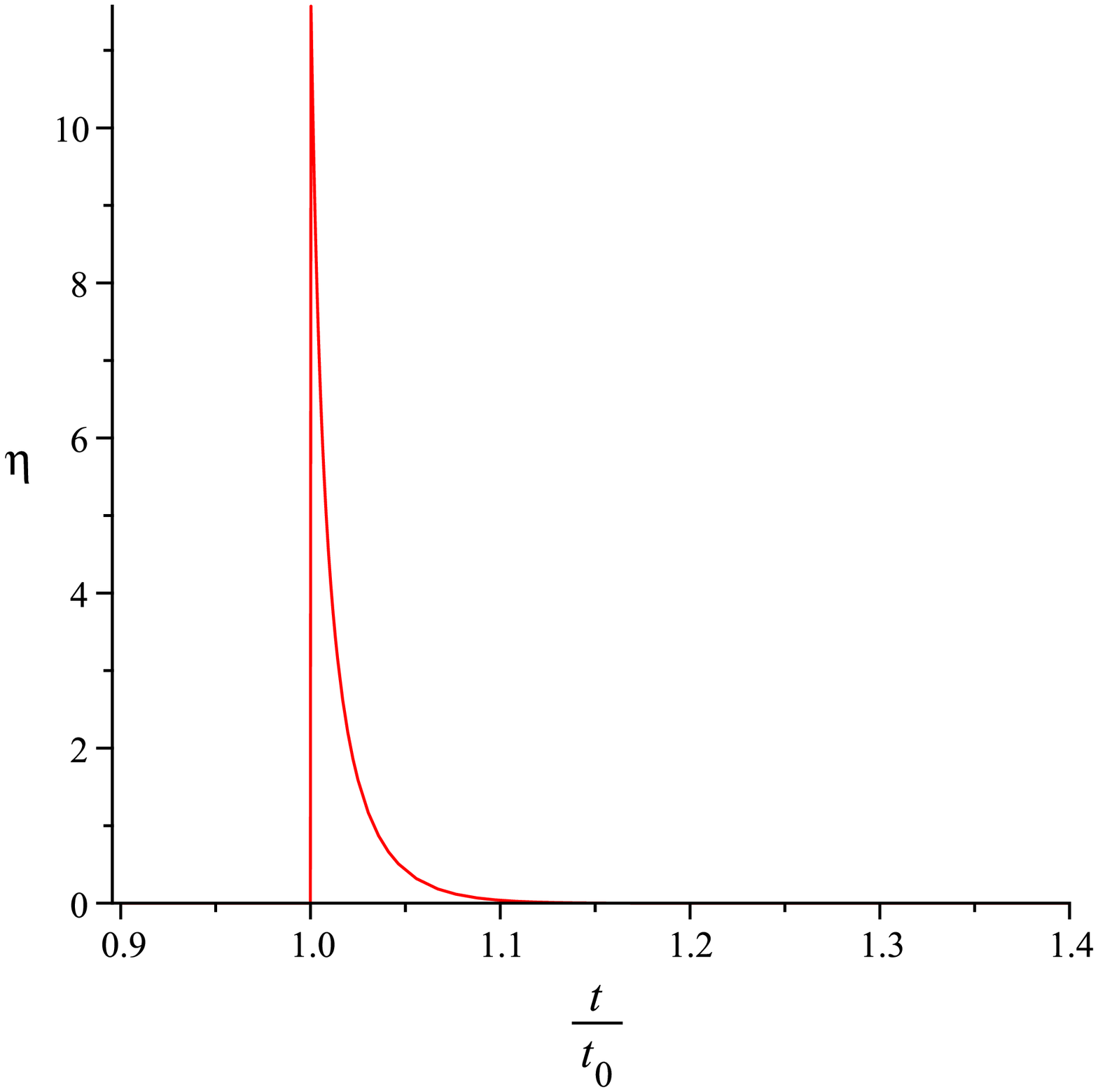}
    \includegraphics*{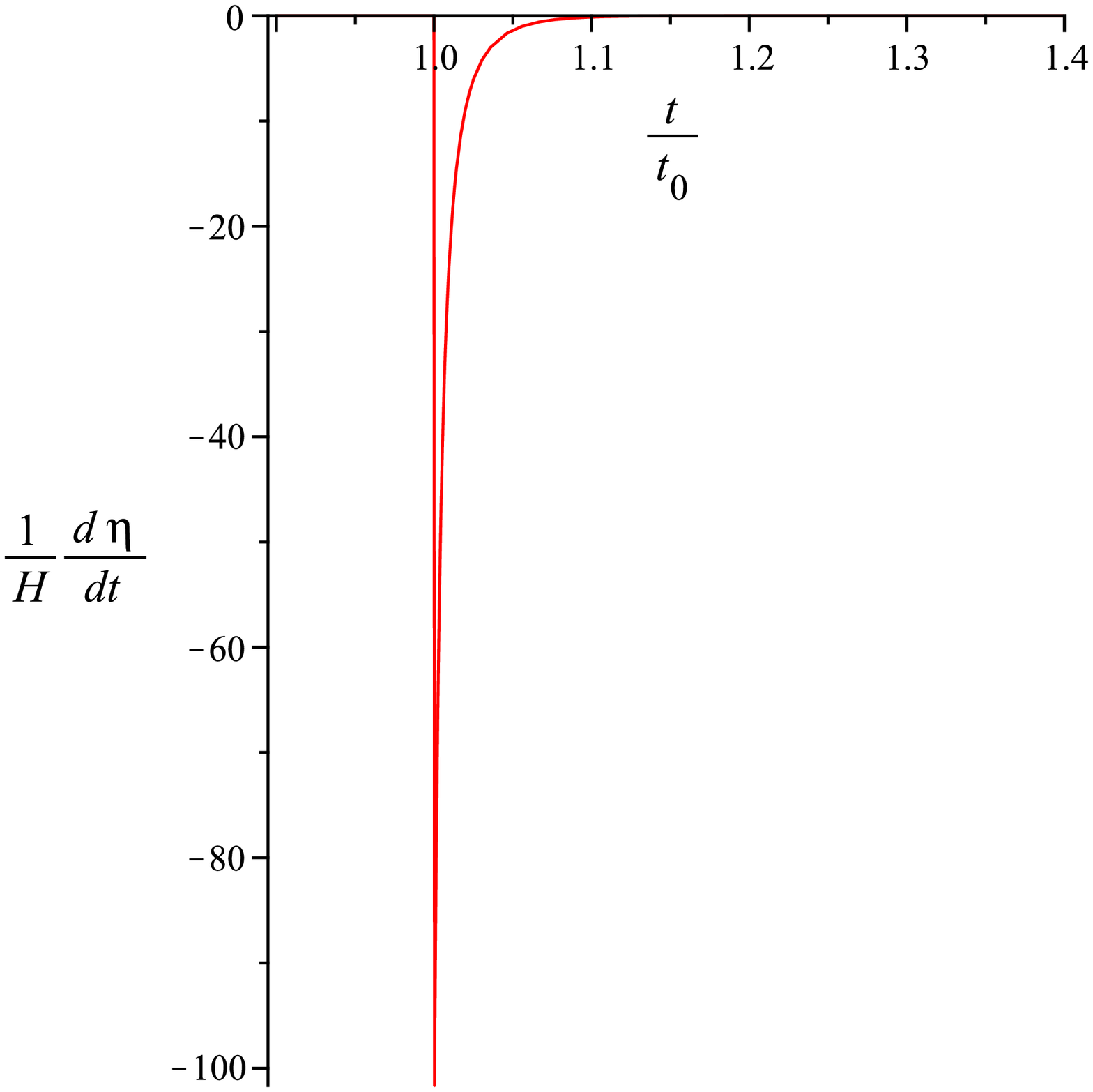}
                 }
 }
\caption{
Plots of $\epsilon$ (left), $\eta$ (center) and $\frac{1}{H}\frac{d\eta}{dt}$ (right) for $A=2$. The heights of the peaks in the central and right plots are approximately $6A$ and $-18A(1+A)$, respectively. While $\epsilon$ remains always small, both $\eta$ and $\frac{1}{H}\frac{d\eta}{dt}$ become large after the transition and will source a large bispectrum.}\label{SRPplots}
\end{figure}

In Fig. \ref{SRPplots}, we plot we slow-roll parameters $\epsilon$, $\eta$ and $\frac{\dot \eta}{H}$ for an amplitude of the transition of $A=2$. We chose $t_0=-1/H\ln H$ (which corresponds to conformal time $\tau_0=-1$). One can see that for our choice of parameters $\epsilon$ is always small, while $\eta$ and $\frac{\dot \eta}{H}$ become large for some time after the transition time $t_0$. During this time the slow-roll expansion breaks down and this will introduce some scale-dependence in the power spectrum and produce non-Gaussian perturbations as we shall see in the next two sections.

%%%%%%%%%%%%%%%%%%%%%%%%%%%%%%%%%%%%%%%%%%%%%%%%%%%%%%%%%%%%%%%%%%%%
%%%%%%%%%%%%%%%%%%%%%%%%%%%%%%%%%%%%%%%%%%%%%%%%%%%%%%%%%%%%%%%%%%%%%
%%%%%%%%%%%%%%%%%%%%%%%%%%%%%%%%%%%%%%%%%%%%%%%%%%%%%%%%%%%%%%%%%%%%%
\section{Perturbations\label{sec:pert}}

In this section, we will discuss linear perturbations of the previous background. We shall present an analytical approximation to the mode function that will allows to obtain an analytical and accurate approximation to the power spectrum on scales sufficiently different from the scale $k_0$ set by the transition. Finally we integrate numerically the equation of motion of the curvature perturbation and obtain the power spectrum which we then compare with the analytical approximation.

On the comoving time-slices, the scalar field fluctuations vanish,
$\delta\phi=0$,
and the three-dimensional spatial metric $h_{ij}$ is perturbed as \cite{Maldacena:2002vr}
\begin{eqnarray}
h_{ij}=a^2e^{2\mathcal{R}}\delta_{ij},
\end{eqnarray}
where tensor perturbations have been neglected because they do not contribute for the tree-level scalar bispectrum and
$\mathcal{R}$ denotes the curvature perturbation on comoving slices.

The linear equation of motion for $\mathcal{R}$ is
\begin{equation}
\frac{\partial}{\partial t}
\left(a^3\epsilon\frac{\partial}{\partial t}\mathcal{R}\right)
-a\,\epsilon\,
\delta^{ij}\frac{\partial^2}{\partial x^i\partial x^j}\mathcal{R}=0\,.
\label{calRceq}
\end{equation}
In Fourier space and using the variable $z=a\sqrt{2\epsilon}$ the previous equation can be written as
\begin{equation}
\mathcal{R}_k''+2\frac{z'}{z}\mathcal{R}_k'+k^2\mathcal{R}_k=0,\label{calRceqFourier}
\end{equation}
where prime denotes derivative with respect to conformal time $\tau$ and $k$ denotes the comoving wavenumber.

Because we assume the potential is vacuum energy dominated, $H$ is constant and the scale factor may be approximated by that of a pure de Sitter universe $a=-1/(H\tau)$ and $-Ht=\ln(-H\tau)$. We choose $a(t=0)=1$.
The inflaton perturbation on flat hypersurfaces $\delta\phi$ is related to $\mathcal{R}$ at first order as
\begin{equation}
\delta\phi=-\frac{\dot\phi}{H}\mathcal{R}.
\end{equation}

The equation of motion for the field perturbation can be written as
\begin{equation}
u''+\left(k^2-\frac{z''}{z}\right)u=0,\label{eqmodefcu}
\end{equation}
where the variable $u$ is defined as $u\equiv a\delta\phi=-\mathrm{sign}(\dot\phi)M_{Pl}z\mathcal{R}$.
At the time of the transition $\tau=\tau_0$ (or $t=t_0$), Eq.(\ref{Klein-Gordon}) implies that $\phi''$ is discontinuous which in turn implies that $z''$ contains a Dirac delta function. Using Eq. (\ref{eqmodefcu}) one can immediately see that also $u''$ contains a Dirac delta function and $u'$ is discontinuous at $\tau=\tau_0$.
One can evaluate this discontinuity as
\begin{equation}
D_0=\lim_{\xi\rightarrow0}\int_{\tau_0-\xi}^{\tau_0+\xi}\frac{z''}{z}d\tau=\lim_{\xi\rightarrow0}\int_{\tau_0-\xi}^{\tau_0+\xi}\frac{\phi'''}{\phi'}d\tau=-m_\phi^2Aa^2(\tau_0)\frac{\phi(\tau_0)}{\phi'(\tau_0)}.
\end{equation}
Thus the matching conditions for $u$ at the point $\tau=\tau_0$ are given by
\begin{equation}
u(\tau_0^+)=u(\tau_0^-), \quad u'(\tau_0^+)=u'(\tau_0^-)+D_0u(\tau_0^-),
\label{matchingconditions}
\end{equation}
where the superscripts $-$ and $+$ mean the lhs and rhs limits at the point $\tau_0$ respectively.
In terms of the comoving curvature perturbation $\mathcal{R}$, these matching conditions imply the continuity of both $\mathcal{R}$ and $\mathcal{R}'$ at $\tau=\tau_0$. This is obviously consistent with the equation of motion for $\mathcal{R}$, Eq. (\ref{calRceqFourier}), in which there is no Dirac delta function. In the next section, it is this last equation (\ref{calRceqFourier}) that we will integrate numerically to find the solution for the mode functions. Regarding the initial conditions we will assume the standard Bunch-Davies vacuum for $\delta\phi$ at $\tau\rightarrow-\infty$. At sufficiently early times, $\tau\rightarrow-\infty$, the mode functions are well approximated by
\begin{equation}
u_{b}=v\equiv\frac{e^{-ik\tau}}{\sqrt{2k}}\left(1-\frac{i}{k\tau}\right),\label{defv}
\end{equation}
where we denote the mode function at $\tau<\tau_0$ by $u_b$. In our numerical analysis, we will use this previous equation to set the initial conditions at the initial integration time $\tau_i$ for each mode $u$ when it is inside the horizon and for $\tau<\tau_0$. Then we change variables to $\mathcal{R}$ and integrate numerically Eq. (\ref{calRceqFourier}).

For long wavelength modes compared with $k_0$, i.e. $k<k_0$, the transition in the mass happens when the modes are outside the horizon and Eq. (\ref{defv}) is a good approximation until the modes cross the horizon. So we will use it to set the initial conditions at some time before horizon crossing and then solve Eq. (\ref{calRceq}) numerically. On the other hand, for modes with wavenumber greater than $k_0$, the sudden change in the mass happens before horizon crossing and we will set the initial conditions using Eq. (\ref{defv}) at the time $\tau=\tau_i<\tau_0$ when it is still a good approximation for the mode function.

To quantize the curvature perturbation, we follow the standard procedure in quantum field theory. We promote $\mathcal{R}$ to an operator that is expanded in terms of creation and annihilation operators and mode functions as
\begin{eqnarray}
\hat{\mathcal{R}}(\tau,\mathbf{k})=\mathcal{R}(\tau,\mathbf{k})a(\mathbf{k})
+\mathcal{R}^*(\tau,-\mathbf{k})a^{\dagger}(-\mathbf{k}),
\end{eqnarray}
where $a$ and $a^\dagger$ satisfy the usual commutation relation
$[a(\mathbf{k}),a^\dagger(\mathbf{k}')]=(2\pi)^3 \delta^{(3)}(\mathbf{k}-\mathbf{k}')$.

The power spectrum of the curvature perturbation is given by
\begin{equation}
2\pi P^{1/2}_\mathcal{R}(k)=\sqrt{2k^3}|\mathcal{R}_k(t_f)|,\label{powerspectrumdef}
\end{equation}
where $t_f$ is the time at which inflation ends. $P_\mathcal{R}(k)$ has been measured observationally to be $P_\mathcal{R}(k_0)=2.95\times10^{-9}A(k_0)$, where $A(k_0)$ is $A(k_0)\sim 0.71-0.75$ and $k_0=0.002\mathrm{Mpc}^{-1}$ \cite{Peiris:2003ff}.

%%%%%%%%%%%%%%%%%%%%%%%%%%%%%%%%%%%%%%%%%%%%%%%%%%%%%%%%%%%%%%%%%%%%%

\subsection{Analytical approximation to the mode function\label{subsec:analyticalapprox}}

In this subsection, we will present an analytical approximation to the mode function. This approximation has been discussed in previous work, see for example \cite{Starobinsky:1992ts,Romano:2008rr}.

For modes with $k<k_0$, the transition happens when the modes are outside the horizon and the only possible time variation of $\mathcal{R}$ is through the discontinuity of $z'/z$ present in Eq. (\ref{calRceqFourier}), however $z'/z=a\dot z/z$ which means that it grows exponentially with time and so the effect of the discontinuity become more and more irrelevant. In fact, Ref. \cite{Leach:2001zf} has shown that the only expected modification with respect to the usual result is for scales around $k=k_0$. Therefore, for $k<k_0$ and for $\tau<\tau_k$, where $\tau_k$ is the horizon crossing time for that mode, we will approximate the mode function by Eq. (\ref{defv}).

For modes with wavenumber $k>k_0$, Eq. (\ref{defv}) is still a good approximation for $\tau<\tau_0$ and we shall use this fact.
For some time after the transition, the slow-roll expansion breaks down and one can not use the slow-rolling solution for the background, however for sufficiently large $k$, the mode is well inside the horizon at the time of the transition and this small violation of slow-roll becomes negligible because the mode equation (\ref{eqmodefcu}) is dominated by the term proportional to $k^2$ and the slow-roll violating term $z''/z$ can be safely neglected. Hence, Eq. (\ref{defv}) is still a good approximation even for $\tau>\tau_0$, including the short non-slow-roll period after the transition. However, the solution for the mode function $u_a$ after $\tau=\tau_0$ is no longer given by just the positive frequency mode function $v$, instead it will be a mixture of positive and negative frequency modes as
\begin{equation}
u_a=\alpha_k v+\beta_k v^*,\label{outgoingmode}
\end{equation}
where the Bogoliubov coefficients $\alpha_k$ and $\beta_k$ can be expressed in terms of $v$, $v^*$ and the outgoing mode function $u_a$ as
\begin{equation}
\alpha_k=-i\left({v^*}' u_a-v^*u_a'\right),\quad \beta_k=i\left(v'u_a-v u_a'\right),\label{Bololiubovcoefficients}
\end{equation}
and they satisfy $|\alpha_k|^2-|\beta_k|^2=1$.
The matching conditions (\ref{matchingconditions}) imply
\begin{equation}
u_a(\tau_0^+)=v_0,\quad u'_a(\tau_0^+)=v'_0+D_0v_0,
\end{equation}
where $v_0=v(\tau_0)$ and together with Eqs. (\ref{Bololiubovcoefficients}) applied at the time $\tau=\tau_0^+$ they can be used to find the Bogoliubov coefficients as
\begin{equation}
\alpha_k=1+iD_0v_0v_0^*=1+i\frac{D_0}{2k}\left(1+\frac{1}{(k\tau_0)^2}\right),\quad \beta_k=-iD_0v_0^2=-i\frac{D_0}{2k}\left(1-\frac{i}{k\tau_0}\right)^2e^{-2ik\tau_0}.\label{Bololiubovcoefficientsexplicit}
\end{equation}

In terms of the original variable, the comoving curvature perturbation, the analytical approximation for the mode function that we use is
\begin{eqnarray}
\mathcal{R}(\tau,k) = \frac{1}{M_{Pl}a(\tau)}
                \left\{
                       \begin{array}{ll}
                          \frac{v(\tau,k)}{\sqrt{2\epsilon(\tau)}} & : k\leq k_0 \,\,\textrm{and}\,\, \tau\leq\tau_k \\
                          \frac{v(\tau,k)}{\sqrt{2\epsilon(\tau_k)}} & : k\leq k_0 \,\,\textrm{and}\,\, \tau>\tau_k \\
                          \frac{v(\tau,k)}{\sqrt{2\epsilon(\tau)}} & : k>k_0 \,\,\textrm{and}\,\, \tau\leq\tau_0\\
                          \frac{\alpha(k)v(\tau,k)+\beta(k)v^*(\tau,k)}{\sqrt{2\epsilon(\tau)}} & : k>k_0 \,\,\textrm{and}\,\, \tau_0<\tau\leq\tau_k\\
                          \frac{\alpha(k)v(\tau,k)+\beta(k)v^*(\tau,k)}{\sqrt{2\epsilon(\tau_k)}} & : k>k_0 \,\,\textrm{and}\,\, \tau>\tau_k
                        \end{array}
                \right.\label{approx}
\end{eqnarray}
where the difference between the last two branches and between the second branch and the first and the third branches is that we fixed the value of the slow-roll parameter $\epsilon$ to its horizon crossing value, i.e. when the modes are super-horizon. We have assumed that $\mathrm{sign}(\dot\phi)=-1$.
Note that $\mathcal{R}(\tau,k)$ is discontinuous on the line $k=k_0$ and for $\tau>\tau_0$. The time derivative is discontinuous on the horizon crossing line $k\tau_k=-1$, in particular the time derivative is discontinuous at the point $(\tau_0,k_0)$, i.e. $\mathcal{R}'(\tau_0,k_0^+)\neq\mathcal{R}'(\tau_0,k_0^-)$, we choose $\mathcal{R}'(\tau_0,k_0)=\mathcal{R}'(\tau_0^-,k_0^+)$. Note that $\mathcal{R}'(\tau_0^-,k_0^\pm)=\mathcal{R}'(\tau_0^+,k_0^\pm)$.
We have solved numerically the mode function equation of motion (\ref{calRceq}) an used the analytical solutions for the background given in section \ref{sec:model} and we found that the previous analytical approximation (\ref{approx}) is a good approximation for the mode functions at any time $\tau$ and for scales far from $k_0$ even if the amplitude of the transition $A$ is of order one. As expected this approximation deteriorates for scales near $k_0$, however for $A$ small this continues to be a good approximation.

%%%%%%%%%%%%%%%%%%%%%%%%%%%%%%%%%%%%%%%%%%%%%%%%%%%%%%%%%%%%%%%%%%%%%

\subsection{The power spectrum of the primordial curvature perturbation\label{subsec:powerspectrum}}

In this subsection, we shall present the result for the power spectrum calculated numerically and compare it with the power spectrum given by the analytical approximation.

As discussed previously, for modes with wavenumber $k$ such that $k\ll k_0$ there can be no super-horizon evolution of $\mathcal{R}$ and the curvature perturbation power spectrum has to be given by the standard formula, which in our present case is
\begin{equation}
P^{1/2}_\mathcal{R}(k)=\frac{H^2}{2\pi|\dot\phi(t_k)|}=\frac{H}{2\pi\lambda^+_b\phi_{0b}}\left(\frac{H}{k}\right)^{\lambda^+_b}\equiv P^{1/2}_<(k),
\label{PSlargescale}
\end{equation}
where $t_k$ is the horizon crossing time, $k=a(t_k)H$. This is the same equation as in the case of no transition. Our analytical approximation consists of using this result all the way down to the scale $k_0$ however for scales near $k_0$ this might not be a good approximation.

For modes with $k>k_0$, using Eq. (\ref{outgoingmode}) in the power spectrum definition (\ref{powerspectrumdef}) one can find that the spectrum is
\begin{equation}
P_\mathcal{R}^{1/2}(k)=\frac{H^2}{2\pi|\dot\phi(t_k)|}|\alpha_k-\beta_k|,
\end{equation}
where the Bogoliubov coefficients are given in Eq. (\ref{Bololiubovcoefficientsexplicit}). In the following plots of our analytical approximation we will use the previous result for scales $k\geq k_0$ but one should keep in mind that in principle this is expected to be a good approximation only for scales $k\gg k_0$.

For $k\gg k_0$, one can find a simple expression for the power spectrum \cite{Romano:2008rr}
\begin{equation}
P_\mathcal{R}^{1/2}(k)=\frac{H^2}{2\pi|\dot\phi(t_k)|}\left[1+\frac{D_0}{k}\left(\sin (2k\tau_0)+\mathcal{O}\left(\frac{1}{k\tau_0}\right)\right)
+\frac{D_0^2}{2k^2}\left(1+\cos (2k\tau_0)+\mathcal{O}\left(\frac{1}{k\tau_0}\right)\right)\right]^{1/2}.
\end{equation}
For later use we define $P^{1/2}_>(k)$ as
\begin{equation}
P^{1/2}_>(k)\equiv \frac{H}{2\pi\lambda^+_a\phi_{0b}}\left(\frac{H}{k}\right)^{\lambda^+_b}\approx\frac{P^{1/2}_<(k)}{1+A}.
\label{powerspectrumasymptoticsmallerscales}
\end{equation}
This is the asymptotic value of the power spectrum for scales much smaller than the transition scale $k_0$. We should note that if $A$ is large then the amplitude of the power spectrum for scales much smaller than $k_0$ is significantly different from the amplitude for scales much larger than $k_0$. The ratio of these amplitudes is proportional to $1+A$.

The spectral index is defined as
\begin{equation}
n_s-1\equiv\frac{d\ln P_\mathcal{R}(k)}{d\ln k}=-2\epsilon-\eta\approx\frac{2}{3}\mu^2+\mathcal{O}(\mu^4)>0,
\end{equation}
where in the last equality we particularized the result for our model and we used the formula for $P_\mathcal{R}$ given on large scales Eq. (\ref{PSlargescale}). This means that for scales larger than $k_0$ the spectral index in the present model is positive. For scales smaller than $k_0$ the power spectrum contains oscillations which are damped. For $k\gg k_0$, the spectral index approaches a constant given by $2/3\mu_a^2$.  Recent observations favor a negative spectral index \cite{Komatsu:2010fb}, however a scale invariant spectrum is still allowed \cite{Parkinson:2010zr}, this constraints the parameters $\mu^2$ and $\mu_a^2$ of the model to be much smaller than one.

\begin{figure}[t]
\centering
 \scalebox{.4}
 {\rotatebox{0}{
    \includegraphics*{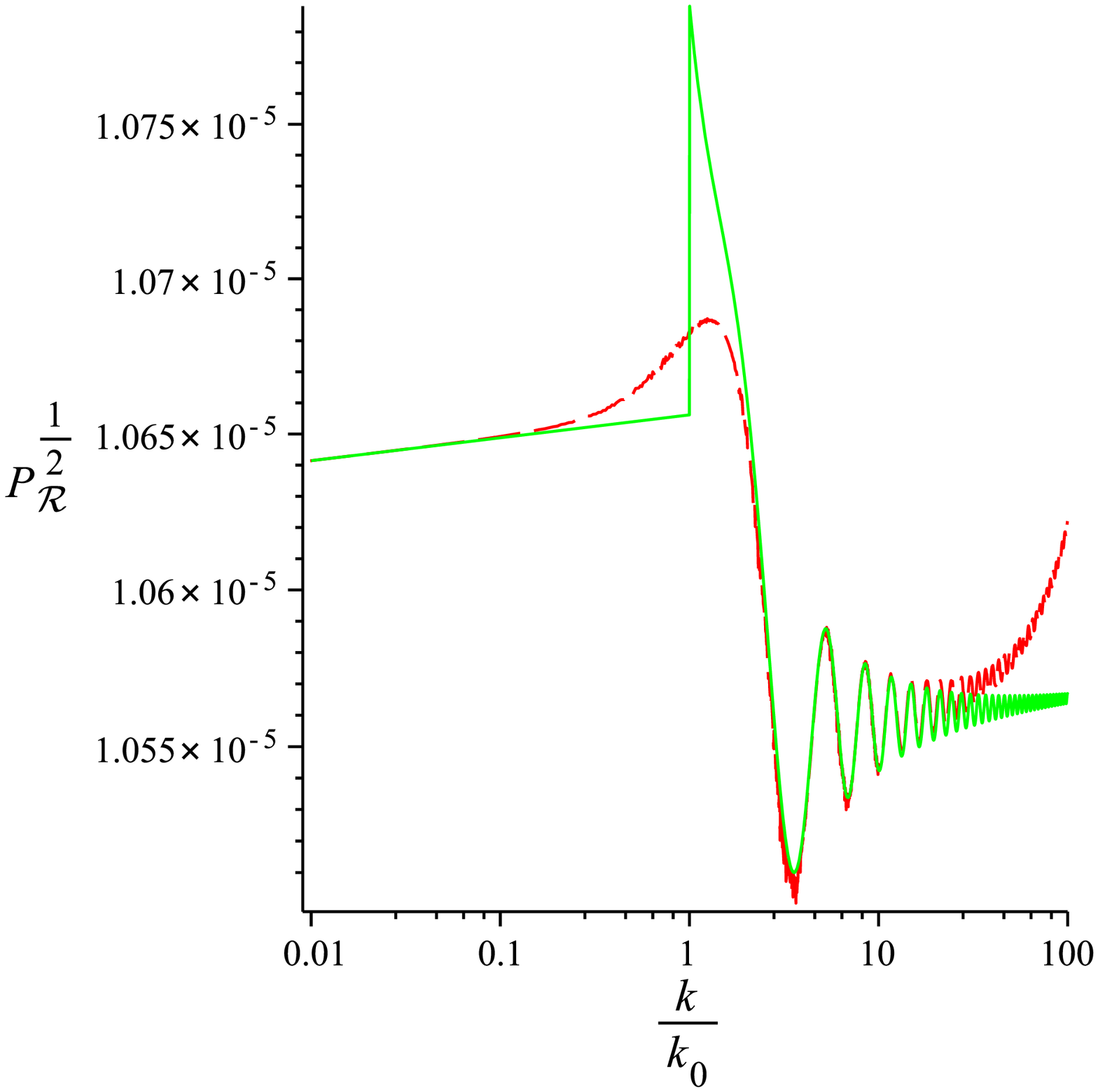}
    \includegraphics*{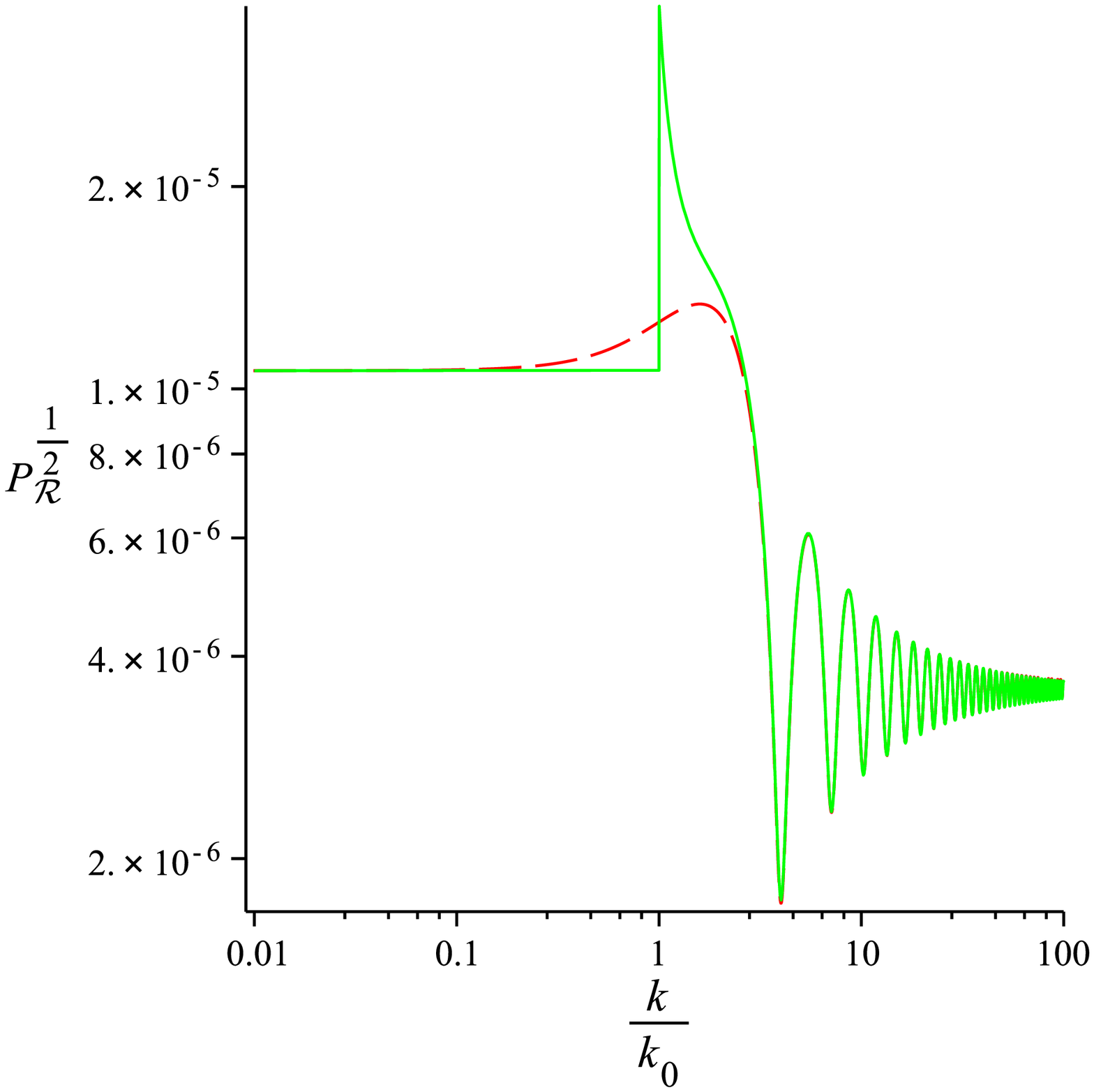}
                 }
 }
\caption{
Plots of $P_\mathcal{R}^{1/2}(k)$ for $A=0.01$ (left) and $A=2$ (right). The red dashed line is the numerical result and the green continuous line is the analytical approximation. It can be seen that the analytical approximation is good for scales different from $k_0$ and in particular it provides a good approximation for the small-scale damped oscillations. On the lhs plot and for $k\gtrsim 10k_0$ the numerical result deviates from the analytical approximation by at most $\mathcal{O}(0.5\%)$, this is due to the numerical error of our integrator. For large $A$ the amplitude of the power spectrum on small scales is significantly different from the amplitude on large scales.}\label{PSplots}
\end{figure}

As can be seen from Fig. \ref{PSplots} the analytical approximation agrees well with the numerical results at large and small scales compared with $k_0$. As expected, this approximation loses some accuracy for modes around $k_0$. This is because these modes leave the horizon around the time of the transition, for them the effect of the violation of the slow-roll conditions is non-negligible and the approximate solution (\ref{defv}) is no longer good.

%%%%%%%%%%%%%%%%%%%%%%%%%%%%%%%%%%%%%%%%%%%%%%%%%%%%%%%%%%%%%%%%%%%%%
%%%%%%%%%%%%%%%%%%%%%%%%%%%%%%%%%%%%%%%%%%%%%%%%%%%%%%%%%%%%%%%%%%%%%%
%%%%%%%%%%%%%%%%%%%%%%%%%%%%%%%%%%%%%%%%%%%%%%%%%%%%%%%%%%%%%%%%%%%%

\section{The bispectrum\label{sec:bispectrum}}

In this section, we will present the calculation of the bispectrum of the primordial curvature perturbation using the so-called in-in formalism \cite{Schwinger:1960qe,Weinberg:2005vy}. We will use the analytical approximation of the mode function described in the previous section and therefore we will be able to obtain analytical expressions for the leading order bispectrum for the present model in certain interesting limits.

In order to use the machinery of the in-in formalism to compute the tree-level three-point correlation function (or bispectrum) one needs to calculate the cubic-order interaction Hamiltonian, see for example \cite{Koyama:2010xj} for a review about this procedure.

The third order action after ignoring many total derivative terms that appear when one simplifies the action by using integrations by parts has been know since the seminal work by Maldacena \cite{Maldacena:2002vr} and can be also found in \cite{Seery:2005wm,Chen:2006nt}, it reads

\begin{eqnarray}
S_3&=&M_{Pl}^2\int dtd^3x\bigg[
a^3\epsilon^2\mathcal{R}\dot{\mathcal{R}}^2+a\epsilon^2\mathcal{R}(\partial\mathcal{R})^2-
2a\epsilon\dot{\mathcal{R}}(\partial
\mathcal{R})(\partial \chi) \nonumber \\ &&\qquad\qquad\qquad +
\frac{a^3\epsilon}{2}\frac{d\eta}{dt}\mathcal{R}^2\dot{\mathcal{R}}
+\frac{\epsilon}{2a}(\partial\mathcal{R})(\partial
\chi) \partial^2 \chi +\frac{\epsilon}{4a}(\partial^2\mathcal{R})(\partial
\chi)^2+ 2 \left(\frac{\eta}{4}\mathcal{R}^2+\tilde{f}(\mathcal{R})\right)\frac{\delta L}{\delta \mathcal{R}}\bigg|_1 \bigg],
\label{cubicaction}
\end{eqnarray}
where we should note that no slow-roll approximation has been made and we define
\begin{eqnarray}
\chi &=& a^2 \epsilon \partial^{-2} \dot{\mathcal{R}}, \qquad
\frac{\delta
L}{\delta\mathcal{R}}\bigg|_1 = a
\left( \frac{d\partial^2\chi}{dt}+H\partial^2\chi
-\epsilon\partial^2\mathcal{R} \right), \\
\tilde{f}(\mathcal{R})&=& \frac{1}{H}\mathcal{R}\dot{\mathcal{R}}+\frac{1}{4a^2H^2}\left[-(\partial\mathcal{R})(\partial\mathcal{R})+\partial^{-2}(\partial_i\partial_j(\partial_i\mathcal{R}\partial_j\mathcal{R}))\right]
+\frac{1}{2a^2H}\left[(\partial\mathcal{R})(\partial\chi)-\partial^{-2}(\partial_i\partial_j(\partial_i\mathcal{R}\partial_j\chi))\right].
\label{redefinition}
\end{eqnarray}
In the previous equations, $\partial^{-2}$ denotes the inverse Laplacian and $\delta
L/\delta\mathcal{R}|_1$ denotes the variation of the quadratic action with
respect to the perturbation $\mathcal{R}$. In order to proceed, usually one performs a field redefinition to eliminate the terms proportional to the linear equations of motion. Recently, it was shown \cite{Arroja:2011yj} that instead of doing field redefinitions one obtains the same result by working with the original variable $\mathcal{R}$ and without ignoring the boundary terms that were omitted in Eq. (\ref{cubicaction}). In the present work we will follow a mixed procedure, i.e. we will use a field redefinition as
\begin{equation}
\mathcal{R}\rightarrow\mathcal{R}_n+\tilde{f}(\mathcal{R}_n)
\end{equation}
to eliminate the last term of the third order action (\ref{cubicaction}) and we will keep the boundary terms as in \cite{Arroja:2011yj} (see also \cite{Burrage:2011hd}).

The previous field redefinition involves terms that contain at least one derivative on $\mathcal{R}$ therefore if one chooses to evaluate them after horizon crossing they should give a negligible contribution to the bispectrum. This implies that at such time the bispectra of $\mathcal{R}$ and $\mathcal{R}_n$ are equal. For this reason from now on we will drop the subscript $n$ and identify $\mathcal{R}_n$ with $\mathcal{R}$.

In Maldacena's calculation \cite{Maldacena:2002vr} and in the following ones \cite{Seery:2005wm,Chen:2006nt} the first three terms in the second line of Eq. (\ref{cubicaction})
which are higher-order in the slow-roll expansion were properly neglected because these authors work at leading order in slow-roll. In the present case, $\epsilon$ is small but $\eta$ and $\eta'$ may be large, this means that the dominant contribution to the three-point function comes from the following terms in the total third order action:
\begin{eqnarray}
S_3&\supset&M_{Pl}^2\int dtd^3x\bigg[
\frac{a^3\epsilon}{2}\frac{d\eta}{dt}\mathcal{R}^2\dot{\mathcal{R}}
+ 2 \frac{\eta}{4}\mathcal{R}^2\frac{\delta L}{\delta \mathcal{R}}\bigg|_1 \bigg]
+M_{Pl}^2\int dtd^3x \frac{d}{dt}\bigg[
-\frac{\eta a}{2}\mathcal{R}^2\partial^2\chi
\bigg],
\end{eqnarray}
where the last total time derivative term comes from the boundary interaction terms recently obtained in \cite{Collins:2011mz} (see also \cite{Arroja:2011yj}).
Instead of using the previous third order action we found it is more convenient to do one integration by parts in time to simplify the action to
\begin{eqnarray}
S_3&\supset&M_{Pl}^2\int dtd^3x\bigg[
-\epsilon\eta a^3\mathcal{R}\dot{\mathcal{R}}^2
-\frac{\epsilon\eta}{2}a\mathcal{R}^2\partial^2\mathcal{R}\bigg].\label{newvertices}
\end{eqnarray}
This form of the action makes it clear that $\mathcal{R}$ is constant outside the horizon to non-linear order as expected \footnote{This procedure is equivalent to the usual one used in for instance Refs. \cite{Chen:2006xjb,Chen:2008wn}.}.

The interaction Hamiltonian in conformal time is
\begin{eqnarray}
H_{int}(\tau) &=&
M_{Pl}^2\int d^3x\bigg[
\epsilon\eta a\mathcal{R}\mathcal{R}'^2
+\frac{\epsilon\eta}{2}a\mathcal{R}^2\partial^2\mathcal{R}\bigg].
\label{Hint3}
\end{eqnarray}

The tree-level three-point correlation function at the time $\tau_e$ after horizon exit is \begin{equation}
\langle\Omega|\hat{\mathcal{R}}(\tau_e,\mathbf{k}_1)\hat{\mathcal{R}}(\tau_e,\mathbf{k}_2)\hat{\mathcal{R}}(\tau_e,\mathbf{k}_3)|\Omega\rangle=
-i\int_{-\infty}^{\tau_e} d\tau a\langle 0|
[
\hat{\mathcal{R}}(\tau_e,\mathbf{k}_1)\hat{\mathcal{R}}(\tau_e,\mathbf{k}_2)\hat{\mathcal{R}}(\tau_e,\mathbf{k}_3),{\hat{H}}_{int}(\tau)]
|0\rangle, \label{interaction}
\end{equation}
where $|\Omega\rangle$ and $|0\rangle$ denotes the interacting vacuum and the free theory vacuum respectively. $[,]$ denotes the standard commutator and the interaction Hamiltonian $\hat{H}_{int}$ is used to evolve the free theory vacuum to the interaction vacuum at the time the three-point function is evaluated \cite{Maldacena:2002vr}.

More explicitly, the bispectrum is
\begin{eqnarray}
\langle\Omega|\hat{\mathcal{R}}(0,\mathbf{k}_1)\hat{\mathcal{R}}(0,\mathbf{k}_2)\hat{\mathcal{R}}(0,\mathbf{k}_3)|\Omega\rangle&\approx&
(2\pi)^3\delta^{(3)}(\mathbf{k}_1+\mathbf{k}_2+\mathbf{k}_3)2M_{Pl}^2\Im\bigg[\mathcal{R}(0,\mathbf{k}_1)\mathcal{R}(0,\mathbf{k}_2)\mathcal{R}(0,\mathbf{k}_3)\int_{\tau_0}^0d\tau\eta\epsilon a^2\mathcal{R}^*(\tau,\mathbf{k}_1)\times
\nonumber\\
&&\left(2\mathcal{R}'^*(\tau,\mathbf{k}_2)\mathcal{R}'^*(\tau,\mathbf{k}_3)-k_1^2\mathcal{R}^*(\tau,\mathbf{k}_2)\mathcal{R}^*(\tau,\mathbf{k}_3)\right)\bigg]+\mathrm{two\,perms.}, \label{bispectrum}
\end{eqnarray}
where we have set $\tau_e=0$ and we evaluate the integral from $\tau_0$ instead of $-\infty$ because for $\tau<\tau_0$, slow-roll is valid, the slow-roll parameters are small and therefore the integral is small. We expect that the approximation of setting $\tau_e=0$ in the integral is good enough because some time after horizon crossing the mode function becomes constant and the integral approximately vanishes. ``two perms." means the other two permutations of $\mathbf{k}_1,\mathbf{k}_2$ and $\mathbf{k}_3$.

\begin{figure}[t]
\centering
 \scalebox{.3}
 {\rotatebox{0}{
    \includegraphics*{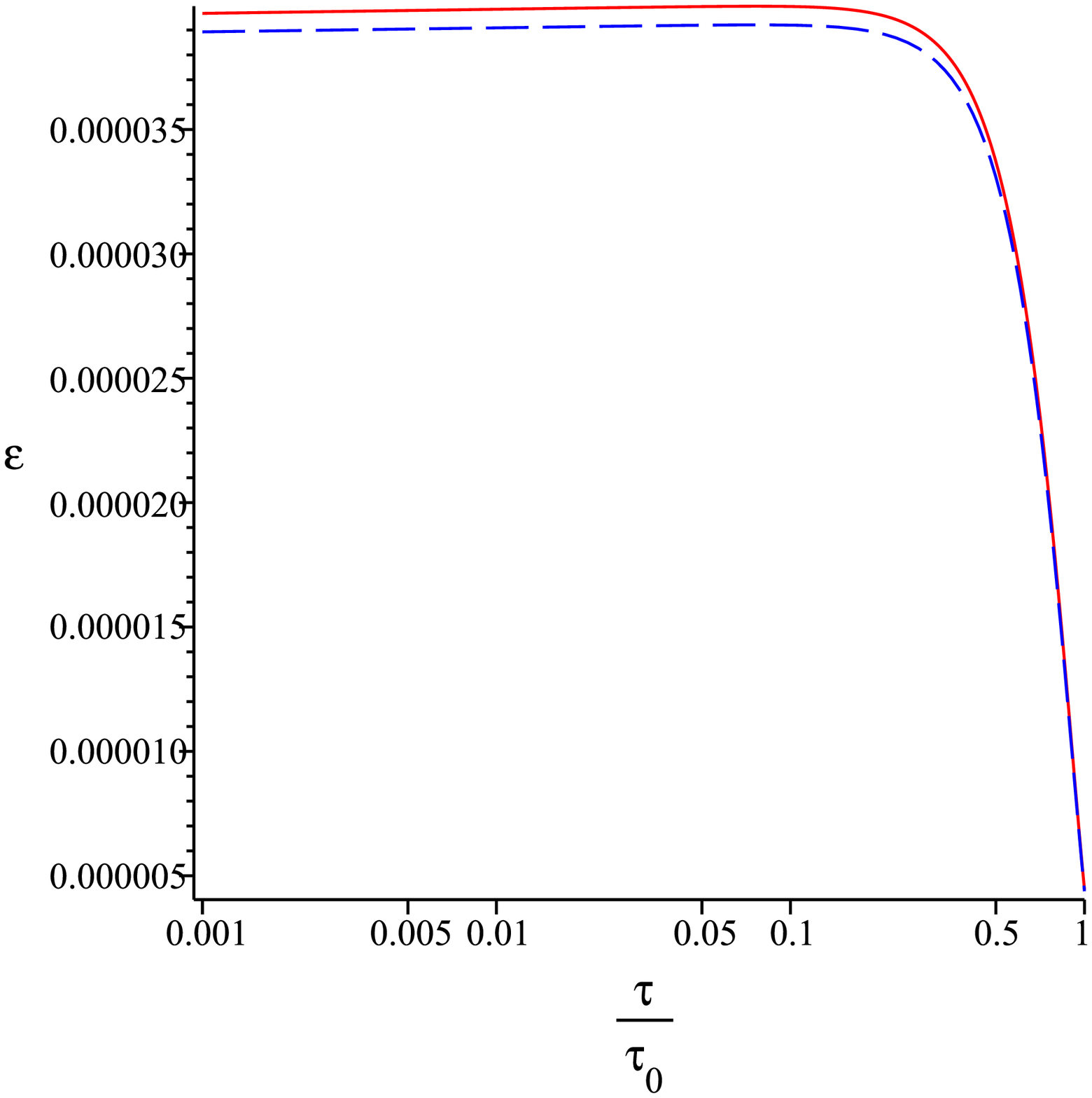}
    \includegraphics*{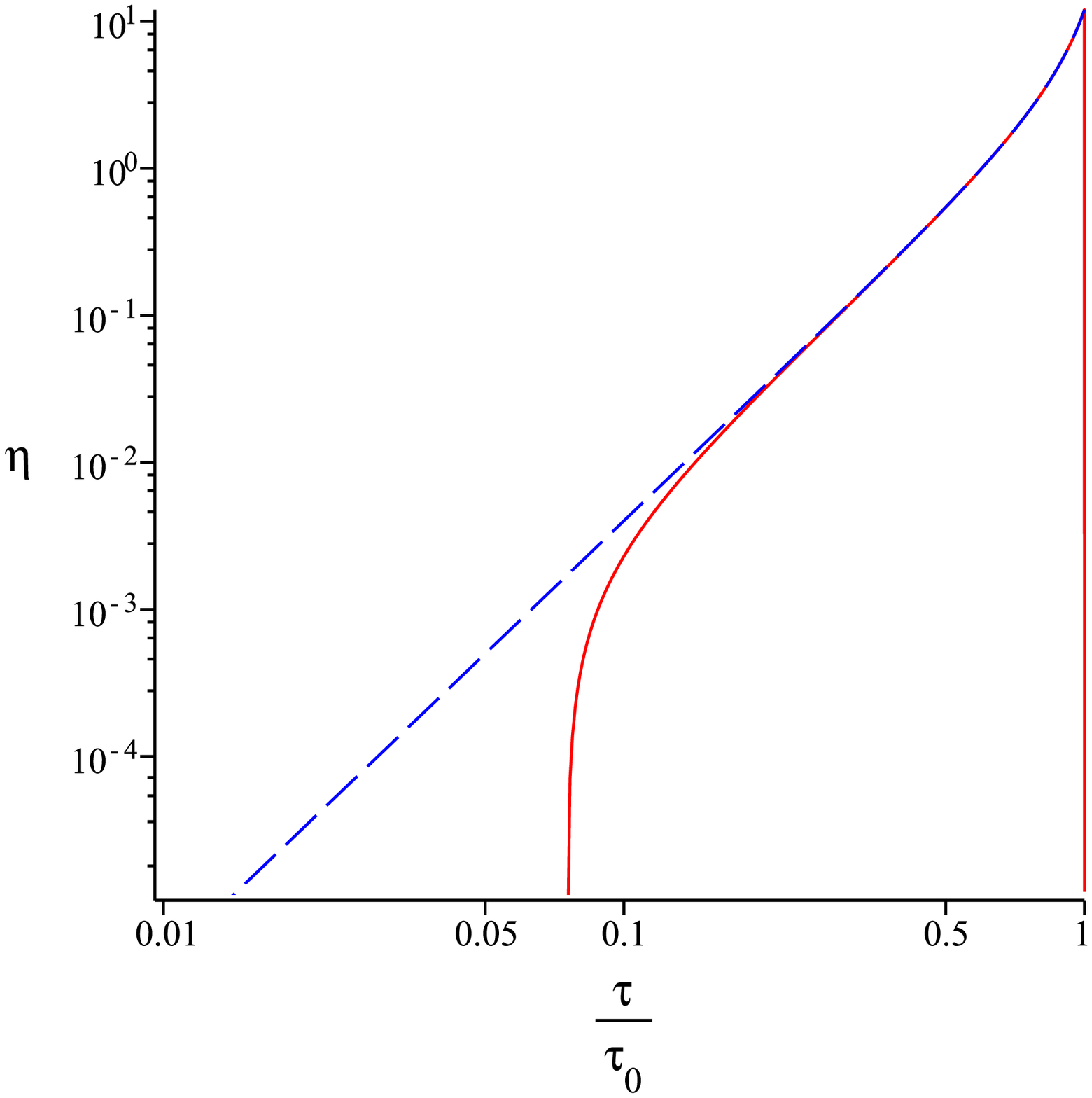}
                 }
 }
\caption{
Plots of $\epsilon$ (left) and $\eta$ (right) for $A=2$. The blue dashed lines are the plots of the approximated expressions Eqs. (\ref{epsilon and eta approx}). The red continuous lines are the plots of the exact analytical solutions found in Sec. \ref{sec:model}. The height of the peak in the rhs plot is approximately $6A$. While $\epsilon$ remains always small, $\eta$ becomes large after the transition and will source a large bispectrum. The simple analytical approximation to $\eta$ becomes poor some time after the transition but from this time $\eta$ is small anyway and we expect a very small error to the bispectrum coming from this deviation.}\label{SRPplots2}
\end{figure}

In the following calculation of the bispectrum, we will approximate the exact analytical solutions for the slow-roll parameters found in Sec. \ref{sec:model} as
\begin{eqnarray}
\epsilon(\tau)\approx\frac{\phi_{0b}^2(\lambda_a^+)^2}{2M_{Pl}^2}a(\tau)^{2\lambda^+_a}\left[1-\frac{A}{1+A}\left(\frac{\tau}{\tau_0}\right)^3\right]^2,
\qquad
\eta(\tau)\approx\frac{-6}{1-\frac{1+A}{A}\left(\frac{\tau_0}{\tau}\right)^3},
\label{epsilon and eta approx}
\end{eqnarray}
which are a good approximation to the exact analytical expressions for times after $\tau_0$, as can be seen in Fig. \ref{SRPplots2}.

Following \cite{Chen:2006xjb}, the meaningful quantity to plot to show the shape of the bispectrum when the three wavenumbers $\mathbf{k}_i,\,i=1,2,3$ have comparable magnitudes is
\begin{equation}
\frac{\mathcal{G}(k_1,k_2,k_3;k_*)}{k_1k_2k_3}=\frac{1}{\delta^{(3)}\left(\mathbf{k}_1+\mathbf{k}_2+\mathbf{k}_3\right)}\frac{(k_1k_2k_3)^2}{(2\pi)^7P_\mathcal{R}^2(k_*)}\langle\Omega|\mathcal{R}(\tau_e,\mathbf{k}_1)\mathcal{R}(\tau_e,\mathbf{k}_2)\mathcal{R}(\tau_e,\mathbf{k}_3)|\Omega\rangle,
\label{Gdividedby3ks}
\end{equation}
where $k_*$ denotes a pivot scale at which the power spectrum $P_\mathcal{R}$ is to be evaluated. In the previous equation one should use the formula for the power spectrum amplitude as given by Eq. (\ref{PSlargescale}) or by Eq. (\ref{powerspectrumasymptoticsmallerscales}) depending on whether one is interested in large or small scales respectively. In this way the oscillations seen in the following plots are truly bispectrum oscillations and are not due to the oscillations already present in the power spectrum. If this quantity is at least of order of a few then there is hope that for instance the Planck satellite will measure it.
In the present work, because the asymptotic amplitudes of the power spectrum for scales much larger and much smaller than the transition scale can be very different if $A$ is large, it makes sense to define two separate functions $\mathcal{G}^<$ and $\mathcal{G}^>$  to study the bispectrum when the three wavenumbers have magnitudes smaller and larger than $k_0$ respectively.  The two new functions are defined as
\begin{equation}
\frac{\mathcal{G}_{<}(k_1,k_2,k_3;k_*)}{k_1k_2k_3}=\frac{1}{\delta^{(3)}\left(\mathbf{k}_1+\mathbf{k}_2+\mathbf{k}_3\right)}\frac{(k_1k_2k_3)^2}{(2\pi)^7P_{<}^2(k_*)}\langle\Omega|\mathcal{R}(\tau_e,\mathbf{k}_1)\mathcal{R}(\tau_e,\mathbf{k}_2)\mathcal{R}(\tau_e,\mathbf{k}_3)|\Omega\rangle,
\label{Gdividedby3kslarger}
\end{equation}
\begin{equation}
\frac{\mathcal{G}_{>}(k_1,k_2,k_3;k_*)}{k_1k_2k_3}=\frac{1}{\delta^{(3)}\left(\mathbf{k}_1+\mathbf{k}_2+\mathbf{k}_3\right)}\frac{(k_1k_2k_3)^2}{(2\pi)^7P_{>}^2(k_*)}\langle\Omega|\mathcal{R}(\tau_e,\mathbf{k}_1)\mathcal{R}(\tau_e,\mathbf{k}_2)\mathcal{R}(\tau_e,\mathbf{k}_3)|\Omega\rangle.
\label{Gdividedby3kssmaller}
\end{equation}

For comparison the so-called local model of the bispectrum is
\begin{equation}
\langle\mathcal{R}(\mathbf{k}_1)\mathcal{R}(\mathbf{k}_2)\mathcal{R}(\mathbf{k}_3)\rangle_{local}=
(2\pi)^7\delta^{(3)}\left(\mathbf{k}_1+\mathbf{k}_2+\mathbf{k}_3\right)\frac{3}{10}f_{NL}P_\mathcal{R}^2\frac{\sum_ik_i^3}{\Pi_ik_i^3},
\end{equation}
where the sign of $f_{NL}$ was chosen to agree with the WMAP team definition \cite{Komatsu:2010fb} but it has the opposite sign of the definition present in \cite{Maldacena:2002vr,Chen:2006xjb} for example. We should stress that the definition of $f_{NL}$ in the present paper is different from the definition in \cite{Komatsu:2010fb} but both definitions agree for a scale invariant power spectrum.

Following the definition of the function $F_{NL}$, which is useful to study the squeezed limit of the bispectrum and was first introduced in Ref. \cite{Hannestad:2009yx}, here and for the reason discussed above, we also define two functions $F_{NL}^<$ and $F_{NL}^>$ as
\begin{equation}
F_{NL}^<(k_1,k_2,k_3;k_*)\equiv\frac{10k_1k_2k_3}{3\sum_ik_i^3}\frac{\mathcal{G}_<(k_1,k_2,k_3;k_*)}{k_1k_2k_3},
\end{equation}
\begin{equation}
F_{NL}^>(k_1,k_2,k_3;k_*)\equiv\frac{10k_1k_2k_3}{3\sum_ik_i^3}\frac{\mathcal{G}_>(k_1,k_2,k_3;k_*)}{k_1k_2k_3},
\end{equation}
which reduce respectively to $\mathcal{G}_<(k_1,k_2,k_3;k_*)/(k_1k_2k_3)$ and $\mathcal{G}_>(k_1,k_2,k_3;k_*)/(k_1k_2k_3)$ in the approximately equilateral case and to $f_{NL}$ for the local model.

%%%%%%%%%%%%%%%%%%%%%%%%%%%%%%%%%%%%%%%%%%%%%%%%%%%%%%%%%%%%%%%%%%%%

\subsection{The squeezed limit\label{subsec:squeezed}}

Using the analytical approximations for the mode function (\ref{approx}) and for the slow-roll parameters (\ref{epsilon and eta approx}) one can calculate the integrals analytically, for scales larger than $k_0$, or numerically for any scale.

\begin{figure}[t]
\centering
 \scalebox{.35}
 {\rotatebox{0}{
    \includegraphics*{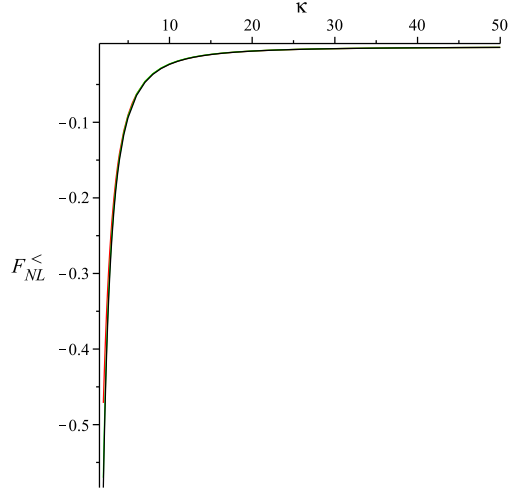}
                 }
 }
\caption{
Plots of $F_{NL}^<(k_0/500,k_0/\kappa,k_0/\kappa;k_0)$ for $A=2$. For the red line, the bispectrum integral in Eq. (\ref{bispectrum}) was calculated numerically and we used the analytical approximation for the mode function. The green and black lines are the plots made using the first and second parts of Eq. (\ref{FNLLargeScales}) respectively. It can be seen that the lines are practically indistinguishable. For scales $k\sim k_0$, i.e. $\kappa$ of order one, our analytical approximation for the mode functions should break down and the result in this plot should not be taken seriously.}\label{FNLLargeScalesFig}
\end{figure}

In the squeezed limit for large scales, i.e. $k_1\ll k_2,k_3\sim k$ and $k\ll k_0$, $F_{NL}^<$ is
\begin{equation}
F_{NL}^<(k_1,k,k;k_*)\approx-\frac{1}{12}\frac{A(8+3A)}{(1+A)^2}\left(\frac{\lambda^+_a}{\lambda^+_b}\right)^2\left(\frac{k}{k_0}\right)^2\left(\frac{a^{\frac{\lambda^+_a}{\lambda^+_b}}(\tau_0)a^2(\tau_*)}{a(\tau_{k_1})a^2(\tau_k)}\right)^{2\lambda^+_b}
\approx-\frac{1}{12}A(8+3A)\left(\frac{k}{k_0}\right)^2. \label{FNLLargeScales}
\end{equation}
The suppression of $F_{NL}$ by the square of the ratio $k/k_0$ has been previously found using the next-order gradient expansion method \cite{Takamizu:2010xy} and using the consistency relation \cite{Ganc:2010ff}.
In this limit and for values of $A$ of order one the quantity $F_{NL}$ is too small as can be seen in Fig. \ref{FNLLargeScalesFig}.

\begin{figure}[t]
\centering
 \scalebox{.7}
 {\rotatebox{0}{
    \includegraphics*{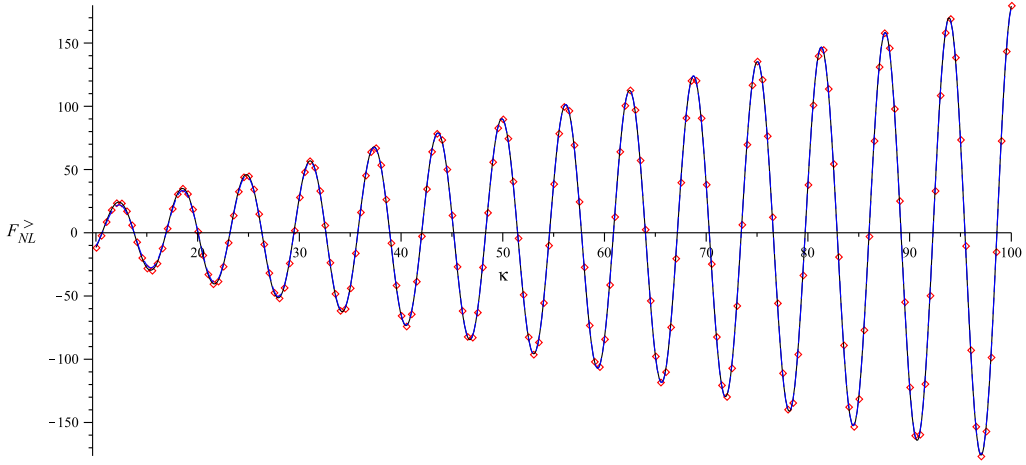}
                 }
 }
\caption{
Plots of $F_{NL}^>(\kappa k_0,1000k_0,1000k_0;k_0)$ for $A=2$. For the red dots, the bispectrum integral in Eq. (\ref{bispectrum}) was calculated numerically and we used the analytical approximation for the mode function. The black line is the plot made using the last line of Eq. (\ref{FNLSmallScales}). The blue dashed line is the plot made using only the first term in the last line of Eq. (\ref{FNLSmallScales}). It can be seen that both lines are almost indistinguishable and are a good approximation for the red points. Both black and blue lines were normalized at the point $\kappa=50$ with the normalization fitting formula Eq. (\ref{F_NL50k01000k0}) and using the first set of constants of Eq. (\ref{F_NL50k01000k0constants}).}
\label{FNLSmallScales1Fig}
\end{figure}

\begin{figure}[t]
\centering
 \scalebox{.8}
 {\rotatebox{0}{
    \includegraphics*{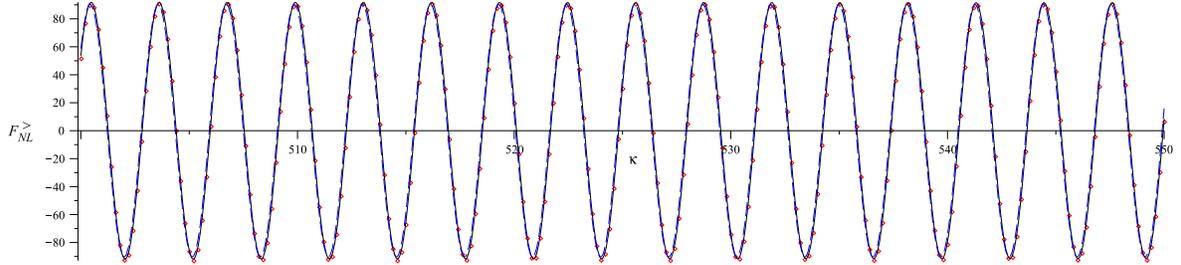}
                 }
 }
\caption{
Plots of $F_{NL}^>(50 k_0,\kappa k_0,\kappa k_0;k_0)$ for $A=2$. For the red dots, the bispectrum integral in Eq. (\ref{bispectrum}) was calculated numerically and we used the analytical approximation for the mode function. The black line is the plot made using the last line of Eq. (\ref{FNLSmallScales}). The blue dashed line is the plot made using only the first term in the last line of Eq. (\ref{FNLSmallScales}). The two lines are almost indistinguishable in the plot and they are good fits to the red points. Both black and blue lines were normalized at the point $\kappa=1000$ with the normalization fitting formula Eq. (\ref{F_NL50k01000k0}) and using the first set of constants of Eq. (\ref{F_NL50k01000k0constants}). }\label{FNLSmallScales2Fig}
\end{figure}

\begin{figure}[t]
\centering
 \scalebox{.35}
 {\rotatebox{0}{
    \includegraphics*{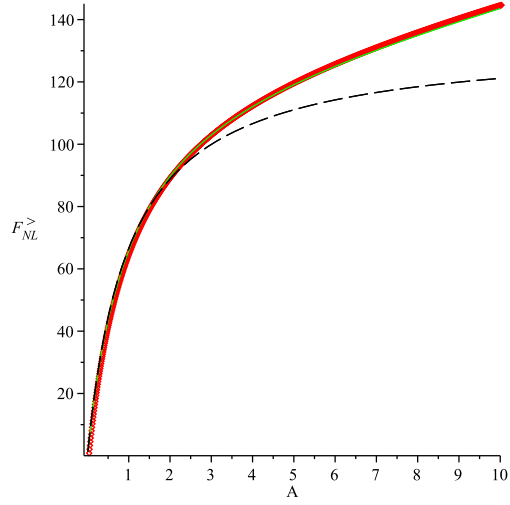}
                 }
 }
\caption{Plots of $F_{NL}^>(50k_0,1000k_0,1000k_0;k_0)$ as a function of $A$. For the red dots, the bispectrum integral in Eq. (\ref{bispectrum}) was calculated numerically and we used the analytical approximation for the mode function. The continuous green line and the black dashed line are the plots made using the fitting formula Eq. (\ref{F_NL50k01000k0}) with the first and second set of constants as in Eq. (\ref{F_NL50k01000k0constants}) respectively. It can be seen that the green line fits well the red points. The line deviates from the red points at most $2\%$.}
\label{F_NL50k01000k0fig}
\end{figure}

To calculate the bispectrum on small scales we used a further approximation in the analytical expressions for the mode function Eq. (\ref{approx}). We fixed the value of $\epsilon(\tau)$ to the horizon crossing value of that mode $\epsilon(\tau_k)$ for any time after and before horizon crossing.
We found for the squeezed limit on small scales, i.e. $k_1\ll k_2,k_3\sim k$ and $k_0\ll k_1$
\begin{eqnarray}
F_{NL}^>(k_1,k,k;k_*)&\approx&\left(\frac{\lambda_b^+}{\lambda_a^+}\right)^4\frac{a^{4\lambda_b^+}(\tau_*)}{a^{6\lambda_b^+}(\tau_0)}\left(\frac{a^4(\tau_0)}{a(\tau_{k_1})a^2(\tau_{k})}\right)^{2\lambda_a^+}
\left(\frac{5A}{4}\right)
\nonumber\\
&&\qquad\times\left[\frac{2k+k_1}{k_0}\frac{k_1}{k}\sin\left(\frac{2k+k_1}{k_0}\right)+(2-3fA)\cos\left(\frac{2k+k_1}{k_0}\right)-3fA\cos\left(\frac{-2k+k_1}{k_0}\right) \right]\nonumber\\
&\approx&\frac{5}{4}\frac{A}{(1+A)^4}\left[\frac{2k+k_1}{k_0}\frac{k_1}{k}\sin\left(\frac{2k+k_1}{k_0}\right)+(2-3fA)\cos\left(\frac{2k+k_1}{k_0}\right)-3fA\cos\left(\frac{-2k+k_1}{k_0}\right) \right],
\nonumber\\\label{FNLSmallScales}
\end{eqnarray}
where the parameter $f$ should be equal to one. If $f=0$ this is equivalent to set the Bogoliubov coefficients to $\alpha_k=1, \beta_k=0$ (the expected behavior in the infinite frequency limit).
We found that the previous additional approximation of fixing the value of $\epsilon(\tau)$ to the horizon crossing value of that mode $\epsilon(\tau_k)$ for any time after and before horizon crossing is not a good approximation in the calculation of the amplitude of the bispectrum, this is because for some time after the transition the value of $\epsilon(\tau)$ varies significantly due to the presence of the background decaying mode of the scalar field. However, as can be seen in Figs. \ref{FNLSmallScales1Fig} and \ref{FNLSmallScales2Fig} this approximation reproduces the shape very well. If one allows $\epsilon(\tau)$ to vary, the amplitude turns out to be larger than in the fixed $\epsilon(\tau)$ case. In the varying $\epsilon(\tau)$ case, we were unable to calculate the bispectrum integral analytically so we proceeded numerically in order to find the amplitude, the result for $A=2$ can be found in Figs. \ref{FNLSmallScales1Fig} and \ref{FNLSmallScales2Fig}.
We were able to find a simple and accurate fitting formula to the numerical result for the amplitude of $F_{NL}^>$ on small scales as
\begin{equation}
F_{NL}^>(50k_0,1000k_0,1000k_0;k_0)\approx\frac{9}{200A(1+A)}C_1A^{C_2}e^{C_3A},
\label{F_NL50k01000k0}
\end{equation}
where the constants are
\begin{equation}
C_1\approx2873.31,\, C_2\approx2.00,\, C_3\approx0.02 \qquad\mathrm{or}\qquad C_1\approx2961.32,\, C_2=2,\, C_3=0.
\label{F_NL50k01000k0constants}
\end{equation}
In the range of values of $A$ as $A=[0.01,10]$, for the first and second set of constants the approximation is good to $2\%$ and $15\%$ respectively. In Fig. \ref{F_NL50k01000k0fig} we plot the value of $F_{NL}^>(50k_0,1000k_0,1000k_0;k_0)$ as a function of $A$ using the numerical result and the above fitting formula (\ref{F_NL50k01000k0}).

As can be seen in Eq. (\ref{FNLSmallScales}) or in Fig. \ref{FNLSmallScales1Fig}, the envelope of the sinusoidal grows with $k_1$, this can have important consequences regarding the magnitude of the non-Gaussianity of this model and its potential observability even for $A$ relatively small. This growth might be worrisome because the bispectrum would increase without bound for very small scales, however one should note that at some scale our toy model, i.e. replacing an otherwise smooth transition by a Heaviside function, ceases to make sense. This scale gives a natural cut-off below which the previous result can not be applied. For instance in Refs. \cite{Elgaroy:2003hp,Romano:2008rr} it was argued that for the particular model under consideration there the Heaviside function approximation was valid for $k<10^3k_0$. For scales smaller than this cut-off and if the width of the transition is taken into account then the linear growth should stop and the amplitude should decay to small values because for these sufficiently high frequencies the transition becomes an adiabatic change of the slow-roll parameters and in this case we expect nearly Gaussian perturbations.

With a simple order of magnitude argument it is possible to estimate the range of wavenumbers $\Delta k$ over which we expect to have a large deviation from Gaussianity as follows. If instead of a sharp transition we have a smooth transition with a width in field space given by $\Delta\phi$ then for the vacuum dominated potential Eq. (\ref{potential}) one can find that the transition happens over a number of e-foldings $\Delta N$ as $\Delta N\sim(3\Delta\phi)/(\mu^2\phi_0)$. Using the fact that $\Delta k/k_0\sim\tau_0/\Delta\tau\sim1/(\Delta N)$ one estimates that the range of scales that can be affected by the transition is $\Delta k\sim k_0/(\Delta N)\sim (\mu^2\phi_0)/(3\Delta\phi)k_0$. One can easily see that for a sharp Heaviside transition $\Delta\phi=0$ then $\Delta k\rightarrow\infty$.

%%%%%%%%%%%%%%%%%%%%%%%%%%%%%%%%%%%%%%%%%%%%%%%%%%%%%%%%%%%%%%%%%%%%

\subsection{The equilateral limit\label{subsec:equilateral}}

In the equilateral limit and on large scales, i.e. if $k\ll k_0$, we find
\begin{eqnarray}
\frac{\mathcal{G}_<(k,k,k;k_*)}{k^3}&\approx&-\frac{27}{160}\frac{A(8+3A)}{(1+A)^2}\left(\frac{\lambda^+_a}{\lambda^+_b}\right)^2\left(\frac{k}{k_0}\right)^2\left(\frac{a^{\frac{\lambda^+_a}{\lambda^+_b}}(\tau_0)a^2(\tau_*)}{a^3(\tau_k)}\right)^{2\lambda^+_b}
\approx-\frac{27}{160}A(8+3A)\left(\frac{k}{k_0}\right)^2.\label{Gdividedby3ksLargeScales}
\end{eqnarray}

\begin{figure}[t]
\centering
 \scalebox{.35}
 {\rotatebox{0}{
    \includegraphics*{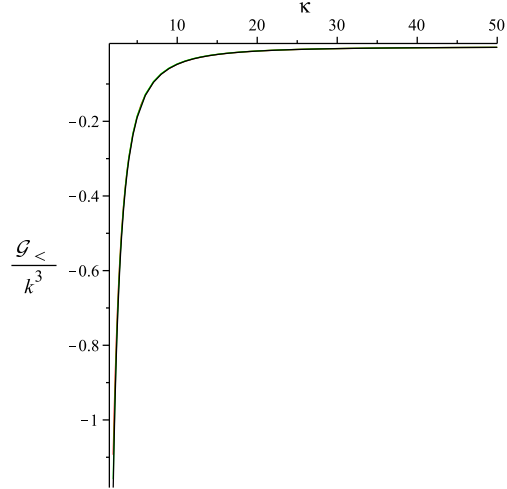}
                 }
 }
\caption{
Plots of $\mathcal{G}_<(k,k,k;k_0)/k^3$ with $k=k_0/\kappa$ for $A=2$. For the red line, the bispectrum integral in Eq. (\ref{bispectrum}) was calculated numerically and we used the analytical approximation for the mode function. The green and black lines are the plots made using the first and second parts of Eq. (\ref{Gdividedby3ksLargeScales}) respectively. It can be seen that the lines are practically indistinguishable. For scales $k\sim k_0$, i.e. $\kappa$ of order one, our analytical approximation for the mode functions should break down and again the result in this plot should not be taken seriously.}\label{Gdividedby3ksLargeScalesFig}
\end{figure}
The plot of $\mathcal{G}_<(k,k,k;k_0)/k^3$ for $A=2$ can be seen in Fig. \ref{Gdividedby3ksLargeScalesFig}. The amplitude of the graph is too small to be interesting observationally. This plot highlights the fact that it is very difficult to generate large non-Gaussianity in single-field models of inflation when the scales of interest are well outside the horizon. This is a direct consequence of the constancy of the curvature perturbation on these scales and of the fact that at horizon crossing the perturbation was very Gaussian because the transition had not happened yet.

In the equilateral limit and on small scales, i.e. if $k_0\ll k$, and using the extra approximation in the analytical expression for the mode function Eq. (\ref{approx}) of fixing the value of $\epsilon(\tau)$ to the horizon crossing value for any time, we find
\begin{eqnarray}
\frac{\mathcal{G}_>(k,k,k;k_*)}{k^3}
&\approx&\left(\frac{\lambda_b^+}{\lambda_a^+}\right)^4\frac{a^{4\lambda_b^+}(\tau_*)}{a^{6\lambda_b^+}(\tau_0)}\left(\frac{a^4(\tau_0)}{a^3(\tau_{k})}\right)^{2\lambda_a^+}\frac{9A}{4}
\left[\frac{k}{k_0}\sin\left(\frac{3k}{k_0}\right)+\frac{6-(2+3f)A}{2}\cos\left(\frac{3k}{k_0}\right)-\frac{9Af}{2}\cos\left(\frac{k}{k_0}\right)\right]
\nonumber\\
&\approx&\frac{9}{4}\frac{A}{(1+A)^4}
\left[\frac{k}{k_0}\sin\left(\frac{3k}{k_0}\right)+\frac{6-(2+3f)A}{2}\cos\left(\frac{3k}{k_0}\right)-\frac{9Af}{2}\cos\left(\frac{k}{k_0}\right)\right],
\label{Gdividedby3ksSmallScales}
\end{eqnarray}
where the parameter $f$ should be equal to one. If $f=0$ this is equivalent to set the Bogoliubov coefficients to $\alpha_k=1, \beta_k=0$ (the expected behavior in the infinite frequency limit). The term with the linear growth in the previous equation, i.e. the term proportional to $k/k_0$, remains even if $f=0$. This means its origin is not from the mixture with the negative frequency modes. In fact, this term is present even in the standard slow-roll scenario if we artificially switch off the interaction for $\tau<\tau_0$, switch it on at $\tau_0$ and perform the integration from $\tau_0$. As argued before, if the finite width of a realistic transition is taken into account then for scales smaller than the scale set by this width we expect the previous result to be unapplicable and the amplitude of $\mathcal{G}_>/k^3$ should decay to small unobservable values.

\begin{figure}[t]
\centering
 \scalebox{.8}
 {\rotatebox{0}{
    \includegraphics*{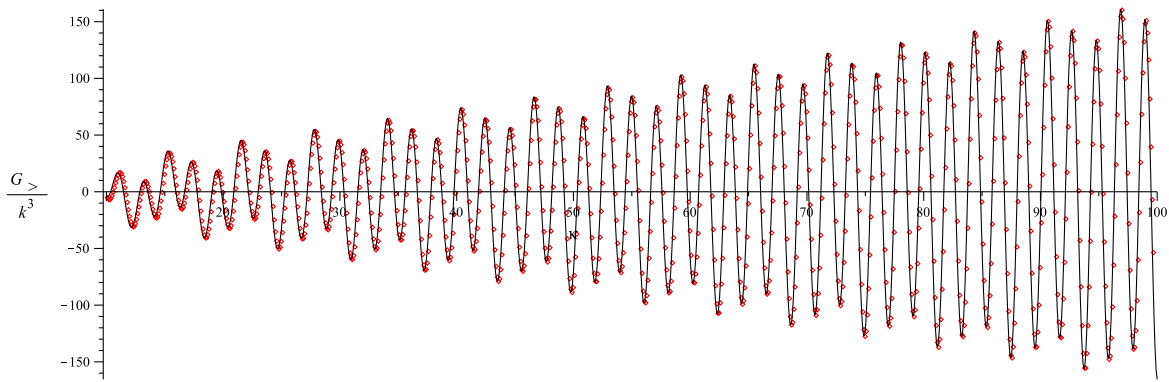}
                 }
 }
\caption{
Plots of $\mathcal{G}_>(k,k,k;k_0)/k^3$ with $k=\kappa k_0$ for $A=2$. For the red dots, the bispectrum integral in Eq. (\ref{bispectrum}) was calculated numerically and we used the analytical approximation for the mode function. The black line is the plot made using the last line of Eq. (\ref{Gdividedby3ksSmallScales}). It can be seen that the black line is a good approximation for the red points. The black line was normalized at the point $\kappa=100$ with the normalization fitting formula Eq. (\ref{Gdividedby3ks100k0}) and using the first set of constants of Eq. (\ref{Gdividedby3ks100k0constants}).}
\label{Gdividedby3ksSmallScalesFig}
\end{figure}

\begin{figure}[t]
\centering
 \scalebox{.4}
 {\rotatebox{0}{
    \includegraphics*{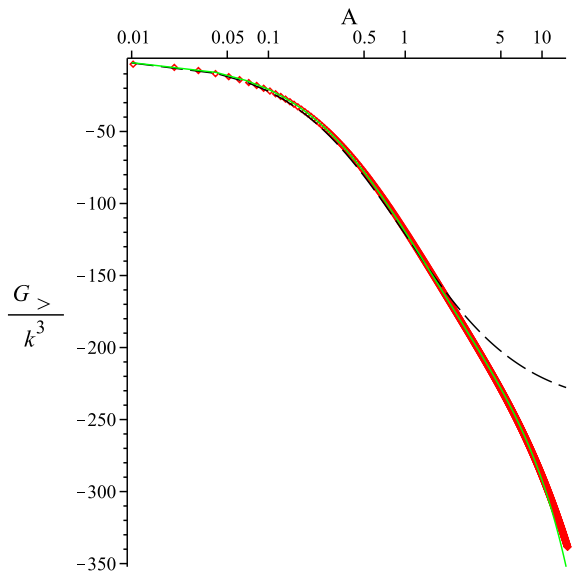}
                 }
 }
\caption{Plots of $\mathcal{G}_>(k,k,k;k_0)/k^3$ with $k=100k_0$ as a function of $A$. For the red dots, the bispectrum integral in Eq. (\ref{bispectrum}) was calculated numerically and we used the analytical approximation for the mode function. The continuous green line and the black dashed line are the plots made using the fitting formula Eq. (\ref{Gdividedby3ks100k0}) with the first and second set of constants as in Eq. (\ref{Gdividedby3ks100k0constants}) respectively. It can be seen that the green line fits well the red points. The line deviates from the red points at most $5\%$.}
\label{Gdividedby3ks100k0fig}
\end{figure}

The plot of the previous expression for $A=2$ is displayed in Fig. \ref{Gdividedby3ksSmallScalesFig}.
As in the squeezed limit, the extra approximation of fixing the value of $\epsilon(\tau)$ to the horizon crossing value is not a good approximation in the calculation of the amplitude of the bispectrum as given by the previous equation. However, to obtain the correct shape that approximation is enough as can be seen in Fig. \ref{Gdividedby3ksSmallScalesFig}.
In order to obtain the correct amplitude of the bispectrum we calculated the integral numerically, but using the analytical approximations for the mode functions, and we were able to find a simple and accurate fitting formula to the numerical result for the amplitude of $\mathcal{G}_>/k^3$ on small scales as
\begin{equation}
\frac{\mathcal{G}_>(k,k,k;k_0)}{k^3}\bigg|_{k=100k_0}\approx\frac{9}{200A(1+A)}\tilde{C}_1A^{\tilde{C}_2}e^{\tilde{C}_3A},
\label{Gdividedby3ks100k0}
\end{equation}
where the constants are
\begin{equation}
\tilde{C}_1\approx-5157.87,\, \tilde{C}_2\approx2.01,\, \tilde{C}_3\approx0.03 \qquad\mathrm{or}\qquad \tilde{C}_1\approx-5399.03,\, \tilde{C}_2=2,\, \tilde{C}_3=0.
\label{Gdividedby3ks100k0constants}
\end{equation}
In the range of values of $A$ as $A=[0.01,15]$, for the first and second set of constants the approximation is good to $5\%$ and $50\%$ respectively. In Fig. \ref{Gdividedby3ks100k0fig} we plot the value of $\mathcal{G}_>/k^3$ with $k=100k_0$ and $k_*=k_0$ as a function of $A$ using the numerical result and the above fitting formula (\ref{Gdividedby3ks100k0}).

It is worth mentioning that in Ref. \cite{Chen:2008wn}, the authors proposed an approximate sharp feature ansatz to be used in data analysis as
\begin{eqnarray}
\mathfrak{f}_{NL}(k)\sin(K/k_0+\mathrm{phase}),
\end{eqnarray}
where $K\equiv k_1+k_2+k_3$ and $\mathfrak{f}_{NL}(k)$ is some envelope. This form seems to fit well with the current model.
Finally, it is important to mention that the linear growth with $k$ that we found in the present model (an Heaviside step function in the inflaton's mass) is in principle observationally distinguishable from the model considered in \cite{Chen:2011zf} where it is the slow-roll parameter $\epsilon$ that has a Heaviside step function discontinuity and the author found a quadratic growth in $K$ \footnote{We would like to thank Xingang Chen for pointing these out to us.}.

%%%%%%%%%%%%%%%%%%%%%%%%%%%%%%%%%%%%%%%%%%%%%%%%%%%%%%%%%%%%%%%%%%%%%%
%%%%%%%%%%%%%%%%%%%%%%%%%%%%%%%%%%%%%%%%%%%%%%%%%%%%%%%%%%%%%%%%%%%%%%
%%%%%%%%%%%%%%%%%%%%%%%%%%%%%%%%%%%%%%%%%%%%%%%%%%%%%%%%%%%%%%%%%%%%%%

\section{Conclusion\label{sec:conclusion}}

In this paper, we introduced an inflationary model driven by a single standard kinetic term scalar field. The field's potential is dominated by the vacuum energy and the mass term changes abruptly at a point. This abrupt change is modeled by an Heaviside step function. This might be seen as a simple toy model for a first order mass phase transition occurring during inflation.

Under the vacuum domination assumption, we have solved the background evolution analytically. As expected for large step sizes and for some time after the transition the slow-roll approximation breaks down.

We have considered linear perturbations and presented an analytical approximation for the mode function that for scales different from the step scale $k_0$ is a good approximation to the numerical mode function for any time. We have computed the power spectrum of the curvature perturbation and showed that for scales smaller than $k_0$ it contains damped oscillations with ``angular frequency" $2/k_0$. This behavior is captured well by our analytical approximation of the spectrum.

Models with temporary violations of slow-roll are expected to produce large non-Gaussianity. For our choice of parameters, we showed that the dominant contribution to the bispectrum comes from the terms containing $\eta$ in the third-order interaction Hamiltonian.
Using the analytical approximation of the mode function, we have computed (analytically when it was possible, otherwise numerically) the bispectrum produced by these terms in certain interesting limits. We analyzed two well known limiting triangle configurations of the three momentum vectors on which the bispectrum depends on.

First, for the equilateral limit, we found that for scales much larger than $k_0$ the quantity $\mathcal{G}_</k^3$ decreases with the square of the ratio $k/k_0$ which implies that for values of $A$ of order one, its value is too small to be of observational interest. For scales much smaller than $k_0$ it oscillates with an ``angular frequency" of approximately $3/k_0$. On these scales $\mathcal{G}^>/k^3$ is linearly enhanced towards large $k$ as $\mathcal{G}_>/k^3\propto k/k_0$. This enhancement may be used to push the value of $\mathcal{G}_>/k^3$ to within observational range even for a small amplitude $A$ of the step. It would be interesting to consider observational constraints on this kind of strongly scale dependent and oscillatory bispectrum models but this is outside the scope of the present paper.

Secondly, we considered the squeezed limit of the bispectrum. For scales much smaller than $k_0$, again we found an oscillatory behavior in $k$ with ``angular frequency" $2/k_0$. When the size of the smallest side of the squeezed triangle is $k_1\gg k_0$ then the quantity $F_{NL}^>$ has as enhancement factor of $k_1/k_0$. This might have important consequences regarding the detectability of the present signal in the data.
For larger scales, we found that the quantity $F_{NL}^<$ is suppressed by the square of the ratio $k/k_0$ which implies that for values of $A$ of order one, its value is too small to be observational relevant.

We have shown that for a wide choice of the model parameter $A$, the squeezed and equilateral limits of the bispectrum can be large for scales much smaller than $k_0$. This represents a significant enhancement with respect to the canonical single field slow-roll model result. In particular, the squeezed limit of the bispectrum can be large in comparison with any single field model of inflation known to the authors. See however Refs. \cite{Agullo:2010ws,Ganc:2011dy} for non-vacuum initial state scenarios which also give significant enhancement factors but which have been shown \cite{Ganc:2011dy} not to be enough to put the CMB bispectrum within the reach of the Planck satellite. It would be very interesting to repeat that analysis for the present model.

The analytical approximation for the mode function used in this paper breaks down for scales close to the scale set by the transition so the present method cannot be applied to compute the bispectrum at these scales. In order to perform that computation one has to numerically integrate the equation of motion for the curvature perturbation and then calculate the bispectrum integral numerically. We leave this for future work. Finally, it would also be interesting to study the trispectrum of this model and this will be presented in a forthcoming publication.

%%%%%%%%%%%%%%%%%%%%%%%%%%%%%%%%%%%%%%%%%%%%%%%%%%%%%%%%%%%%%%%%%%%%%%%
%%%%%%%%%%%%%%%%%%%%%%%%%%%%%%%%%%%%%%%%%%%%%%%%%%%%%%%%%%%%%%%%%%%%%%%%
%%%%%%%%%%%%%%%%%%%%%%%%%%%%%%%%%%%%%%%%%%%%%%%%%%%%%%%%%%%%%%%%%%%%%%%%

\begin{acknowledgments}

We would like to thank Xingang Chen, Kazuya Koyama and Takahiro Tanaka for interesting discussions and useful comments on an early version of the manuscript. FA is also thankful to the organizers of the workshop ``The Cosmological Perturbation and the Cosmic Microwave Background", March 2011 at the Yukawa Institute for Theoretical Physics, Kyoto University for their invitation and financial support during which part of this work was done. Finally, he thanks the Asia Pacific Center for Theoretical Physics for its hospitality and support during the $2^{nd}$ IEU-APCTP Focus Program on Cosmology and Fundamental Physics, June 2011.
FA acknowledges the support by the World Class University grant no. R32-10130 through the National Research Foundation, Ministry of Education, Science and Technology of Korea. AER is supported by the Taiwan NSC under Project No.\ NSC97-2112-M-002-026-MY3, by Taiwan's National Center for Theoretical Sciences (NCTS). MS is supported in part by JSPS Grant-in-Aid for Scientific Research (A) No.~21244033, and by JSPS Grant-in-Aid for Creative Scientific Research No.~19GS0219. This work was also supported in part by MEXT Grant-in-Aid for the global COE program at Kyoto University, ``The Next Generation of Physics, Spun from Universality and Emergence" and by Korea Institute for Advanced Study under the KIAS Scholar program.
\end{acknowledgments}

%%%%%%%%%%%%%%%%%%%%%%%%%%%%%%%%%%%%%%%%%%%%%%%%%%%%%%%%%%%%%%%%%%%%%%%%%%%%%%%%%%%%%%%%%%%%%%%%%%%%%%%%%%%%%%%%%%%%%%%%%%%
%%%%%%%%%%%%%%%%%%%%%%%%%%%%%%%%%%%%%%%%%%%%%%%%%%%%%%%%%%%%%%%%%%%%%%%%%%%%%%%%%%%%%%%%%%%%%%%%%%%%%%%%%%%%%%%%%%%%%%%%%%%

%\bibliography{bibliography}
%\bibliographystyle{bibstylefile}

\end{document}